\documentclass[linenumbers,preprint,tightenlines,preprintnumbers,aps,eqsecnum,nofootinbib,showpacs,floatfix]{revtex4}

\usepackage{amsmath,amssymb,amsbsy,amsfonts,bm}
\usepackage{graphicx}
\usepackage{capt-of}
\usepackage[mathscr]{euscript}
\usepackage{color}
\usepackage{rotating}
\usepackage{slashed}
\usepackage{float}
\usepackage[colorlinks=true,linkcolor=blue,filecolor=blue,urlcolor=blue,citecolor=blue,pdftex=true,plainpages=false]{hyperref}

\usepackage{setspace}
\graphicspath{{./figures/}}

\newcommand{\order}{{\it O}}
\newcommand{\ba}{\begin{eqnarray}}
\newcommand{\ea}{\end{eqnarray}}
\newcommand{\be} {\begin{equation}}
\newcommand{\ee} {\end{equation}}

\newcommand{\cpt}{\raise0.4ex\hbox{$\chi$}PT}
\newcommand{\scpt}{S\raise0.4ex\hbox{$\chi$}PT}
\newcommand{\rscpt}{rS\raise0.4ex\hbox{$\chi$}PT}

\newcommand{\cM}{\ensuremath{\mathcal{M}}}
\newcommand{\cMI}{\ensuremath{\mathcal{M}_I}}
\newcommand{\cMV}{\ensuremath{\mathcal{M}_V}}

\mathchardef\mhyphen="2D
\newcommand{\trt}{\ensuremath{\textrm{tr}_{\rm{t}}}}

\def\mssq#1{${\rm m/s}^2$}

\def\inlinetilde{\lower0.8ex\hbox{$\,\widetilde{}\,$}}
\def\chpt{\raise0.4ex\hbox{$\chi$}PT}
\def\schpt{S\raise0.4ex\hbox{$\chi$}PT}
\def\rschpt{rS\raise0.4ex\hbox{$\chi$}PT}
\def\figref#1{Fig.~\ref{fig:#1}}

\def\secref#1{Sec.~\ref{sec:#1}}

\def\midtild{\lower1ex\hbox{$\tilde{}$}}

\def\leftvec{{\raise1.5ex\hbox{$\leftarrow$}\kern-.85em}}
\def\leftrightvec{{\raise1.5ex\hbox{$\leftrightarrow$}\kern-.85em}}
\def\half{{\scriptstyle \raise.2ex\hbox{$\frac{1}{2}$}}}
\def\threehalves{{\scriptstyle \raise.15ex\hbox{$\frac{3}{2}$}}}
\def\third{{\scriptstyle \raise.15ex\hbox{$\frac{1}{3}$}}}
\def\twothirds{{\scriptstyle \raise.15ex\hbox{$\frac{2}{3}$}}}
\def\fourth{{\scriptstyle \raise.15ex\hbox{$\frac{1}{4}$}}}

\def\gtwid{{\,\raise.3ex\hbox{$>$\kern-.75em\lower1ex\hbox{$\sim$}}\,}}
\def\ltwid{{\,\raise.3ex\hbox{$<$\kern-.75em\lower1ex\hbox{$\sim$}}\,}}

\def\Tr{\rm{Tr}}
\def\tr{\rm{tr}}
\def\circle{{\Large{\raise-0.15ex\hbox{$\circ$}}}}
\def\sqar{{\raise-0.1ex\hbox{$\Box$}}}   

\def\ie{{\it i.e.},\ }
\def\eg{{\it e.g.},\ }
\def\et{{\it et al.}}
\def\etc{{\it etc.}\ }

\def\cA{{\cal A}}
\def\cB{{\cal B}}
\def\cC{{\cal C}}
\def\cD{{\cal D}}

\def\cL{{\cal L}}
\def\cM{{\cal M}}

\def\cO{{\cal O}}

\def\cU{{\cal U}}
\def\cV{{\cal V}}

\def\rcite#1{Ref.~\cite{#1}}

\def\eqn#1{\label{eq:#1}}

\def\eq#1{Eq.~(\ref{eq:#1})}

\def\eqs#1#2{Eqs.~(\ref{eq:#1}) and (\ref{eq:#2})}
\def\eqsthree#1#2#3{Eqs.~(\ref{eq:#1}), (\ref{eq:#2}) and (\ref{eq:#3})}

\def\nsection#1 #2{\leftline{\rlap{#1}\indent\relax #2}}

\def\prd#1{Phys.\ Rev.\ D {\bf #1}}

\newcommand{\bhp}{{\bf h}${\bm{}'}$}

\newcommand{\bhpp}{{\bf h}${\bm{}''}$}

\begin{document}

\singlespacing

\preprint{LU TP 13-42}

\title{Semileptonic Kaon Decay in Staggered Chiral Perturbation Theory}

\author{C.~Bernard}
\affiliation{Department of Physics, Washington University, St.~Louis, Missouri, USA}

\author{J.~Bijnens}
\affiliation{Department of Astronomy and Theoretical Physics, Lund University, Lund, Sweden}

\author{E.~G\'amiz}
\affiliation{CAFPE and Departamento de F\'{\i}sica Te\'orica y del Cosmos,
Universidad de Granada, Granada, Spain}

\collaboration{Fermilab Lattice and MILC Collaborations}
\noaffiliation

\date{\today}

\begin{abstract}

The determination of $\vert V_{us}\vert$ from kaon semileptonic decays requires the value 
of the form factor $f_+(q^2=0)$, which can be calculated precisely on the lattice. We provide  
the one-loop partially quenched staggered chiral perturbation theory expressions that may be employed to  
analyze staggered simulations 
of $f_+(q^2)$ with three light flavors.
We consider both the case of a mixed action, where the valence and sea sectors have different staggered
actions, and the standard case where these actions are the same. 
The momentum transfer $q^2$ of the form factor is allowed to have an arbitrary value.  
We give results for  the generic situation where the $u$, $d$, and
$s$ quark masses are all different, $N_f=1+1+1$, and for the isospin limit,
$N_f=2+1$.  The expression we obtain for $f_+(q^2)$ is independent of the mass of the
(valence) spectator quark.  In the limit of vanishing lattice spacing, 
our results reduce to the one-loop continuum partially quenched expression for $f_+(q^2)$,
which has not previously been reported in the literature for the  $N_f=1+1+1$ case.
Our expressions have already been used in  staggered lattice analyses of $f_+(0)$,
and should prove useful in future calculations as well.

\end{abstract}


\maketitle

\section{Introduction}

\label{introduccion}

Elements of the Cabibbo-Kobayashi-Maskawa 
(CKM) quark-mixing matrix are fundamental parameters of the weak interactions.
In the Standard Model (SM) of particle physics, the matrix is unitary, so any violation of
unitarity would point to new physical phenomena beyond the SM. 
No evidence of such new physics (NP) 
has yet been observed, but precision tests of unitarity provide stringent constraints on the allowed 
non-standard phenomena and the scale at which they may 
occur~\cite{Cirigliano10}. In particular, tests of the first row of the CKM 
matrix provide bounds in the scale of the NP that can contribute to these 
processes at the same level as those from $Z$-pole 
measurements~\cite{Cirigliano11}. 

The precision that can be achieved in these tests of the first row 
depend on the uncertainty in the determination of $\vert V_{ud}\vert$ 
and $\vert V_{us}\vert$, since $\vert V_{ub}\vert$ is negligible at the 
current level of precision. $\vert V_{ud}\vert$ is extracted from nuclear 
$\beta$ decays \cite{Vud08}, while the most precise determinations of 
$\vert V_{us}\vert$ come from kaon leptonic and semileptonic decays. 
Extracting $\vert V_{us}\vert$ from hadronic $\tau$ decays have 
the potential of being competitive with the above
determinations~\cite{tauVus}, but they are currently limited by uncertainties in the
experimental data~\cite{tauHFAG}. Determinations of $\vert V_{us}\vert$ from 
kaon leptonic and semileptonic decays require non-perturbative inputs 
calculated on the lattice: the ratio of decay constants $f_K/f_\pi$~\cite{Follana:2007uv,Durr:2010hr,Aoki:2010dy,Bazavov:2010hj,Laiho:2011np,Bazavov:2013vwa,Dowdall:2013rya} and the vector form factor 
$f_+(q^2=0)$~\cite{ETMC09,RBC10,Kaneko:2012cta,Bazavov:2012cd,Boyle:2013gsa,KtopiLat2013,KtopiHISQ}, 
respectively. 

The vector form factor $f_{+}(q^2)$ is defined from the hadronic matrix element of a 
vector current for $K\to\pi l\nu$ processes
\ba\label{eq:formfacdef}
\langle \pi (p_\pi)\vert V^\mu \vert K(p_K)\rangle =
f_+(q^2) \left[p_K^\mu + p_\pi^\mu - \frac{m_K^2-m_\pi^2}{q^2}
q^\mu\right]+f_0(q^2)\frac{m_K^2-m_\pi^2}{q^2}q^\mu\,,
\ea
where $q=p_K-p_\pi$ is the momentum transfer and $V^\mu=\bar s \gamma^\mu u$
is the appropriate flavor changing vector current. The CKM matrix element 
$\vert V_{us}\vert$ can thus be extracted using the form factor at zero momentum transfer, 
$f_+(0)$, experimental data for $\Gamma_{K_{l3 (\gamma)}}$, and the relation~\cite{Cirigliano11}
\ba\label{eq:Kl3def}
\Gamma_{K_{l3 (\gamma)}} = \frac{G_F^2M_K^5 C_K^2}{128\pi^3}S_{{\rm EW}}
\vert V_{us}f_+^{K^0\pi^-}(0)\vert^2 I_{Kl}^{(0)} 
\left(1+\delta_{{\rm EM}}^{Kl} + \delta_{{\rm SU(2)}}^{K\pi}\right)\,,
\ea
where the Clebsch-Gordon coefficient $C_K$ is equal to $1$ or $1/\sqrt{2}$
for neutral and charged kaons, respectively; $S_{{\rm EW}}=1.0223(5)$ is the short-distance
universal electroweak correction; and, $I_{Kl}^{(0)}$ is a phase space
integral which depends on the shape of the form factors $f_{\pm}$.
The parameters $\delta_{{\rm EM}}^{Kl}$ and $\delta_{{\rm SU(2)}}^{K\pi}$ contain long-distance
electromagnetic and strong isospin-breaking corrections, respectively~\cite{Cirigliano11}.

The error in $f_+(0)$ from lattice-QCD is now small enough so the uncertainty from 
$\vert V_{us}\vert$ from semileptonic decays in the unitarity test is comparable to that 
from $\vert V_{ud}\vert$~\cite{KtopiHISQ}. Further improvement in the calculation of 
$f_+(0)$ is however necessary in order to reach the same level of precision as the 
experimental input, $\Gamma_{K_{l3 (\gamma)}}$. A key element for the reduction of the error 
in the state-of-the-art calculation of $f_+(0)$~\cite{KtopiHISQ} and previous 
lattice calculations using staggered fermions~\cite{Bazavov:2012cd}, as well as for 
future improvements planed by the Fermilab Lattice/MILC Collaboration, is the use of 
staggered chiral perturbation theory (\schpt) to analyze both the chiral behaviour and 
discretization corrections of the form factor.

Two different types of staggered simulations have been and are being used to determine
$f_+(0)$ on the lattice.  In \rcite{Bazavov:2012cd}, a ``mixed-action'' setup is used:
the valence quarks have the HISQ action~\cite{HISQ}, while the sea quarks have the asqtad 
action~\cite{RMP}.  On the other hand, the second-generation calculation of $f_+(0)$ by 
the Fermilab Lattice/MILC Collaboration~\cite{KtopiLat2013,KtopiHISQ} uses valence HISQ 
quarks on HISQ sea-quark ensembles generated by the MILC Collaboration \cite{HISQ-Ensembles}, 
so there is no mismatch between sea and valence actions: the action is ``unmixed.''  

In this paper we calculate $f_+(q^2)$ in \schpt\   for 
both situations.  The chiral theory in the unmixed case is standard, so we simply review 
the formulation and notation, and then proceed to the calculation of  $f_+(q^2)$.  
The mixed-action case, however,  requires some modifications to the corresponding chiral 
theory, so we work those out  first. It is then straightforward to modify the results of 
the result for to form factor in the unmixed case to take into account the complications 
due to the mixed action.

Staggered quarks have a four-fold multiplicity of ``taste'' degrees of freedom, which result from the
fermion doubling in the discretization of the Dirac equation.
In staggered simulations, the unwanted tastes are removed by taking the fourth root of the quark
determinant.  At nonzero lattice spacing, the rooted theory then suffers from nonlocal violations of unitarity
 \cite{Prelovsek05,BGS06}. However, theoretical arguments 
\cite{Shamir04,Bernard06,Shamir06,BGS08},
as well as other analytical and numerical
evidence \cite{SharpePoS06,KronfeldPoS07,GoltermanPoS08,RMP,Donaldetal11},
indicate that standard QCD, both local and unitary, is recovered in the continuum
limit.  In the chiral theory,  taking rooting into account is straightforward: 
each sea quark loop needs to be multiplied by a factor of $1/4$
\cite{SCHPT}.  This can be accomplished  
systematically by replicating the sea quarks $n_r$
times and taking $n_r=1/4$ in the result of the chiral calculation \cite{Bernard06,BGS08}.
However, it is often easier to use the
quark flow approach \cite{QUARK-FLOW} to locate the loops, and then to simply insert the factors
of $1/4$ by hand. Since the quark-flow approach is also useful for other reasons in our
calculations, we will in general use that method below.

This paper is organized as follows. General features of the form factors for $K_{\ell3}$ decay,
including the Ademollo-Gatto theorem \cite{Ademollo:1964sr},
are discussed in \secref{form}.  In Sec.~\ref{sec:preliminaries} we review the basics
of \schpt. 
Section \ref{sec:mixed} then discusses some details of the chiral perturbation theory
for the mixed-action case,  in which the valence and sea actions are different
versions of staggered quarks, \eg HISQ and asqtad, respectively.  
The 
one-loop chiral calculation of the form factor $f_+(q^2)$ is performed in
\secref{one-loop}.  {Although we consider arbitrary values of the valence quark
masses, the result turns out to be independent of the mass of the valence spectator quark. This is
a special property of $f_+(q^2)$, which must satisfy the Ademollo-Gatto theorem when the active (nonspectator)
valence quarks are degenerate, and would not be true of the form factor $f_-(q^2)$. Both form factors
do depend on the masses of all the quarks in the sea, which must enter symmetrically.
The corresponding mixed-action results are presented in \secref{mixed-results}.
We discuss our results and conclusions in Sec.~\ref{sec:conclusion}. Appendix 
\ref{app:integrals} introduces the needed one-loop momentum integrals and their evaluations, 
while Appendix \ref{ref:appB} collects formulas in the special case of exact isospin
in the sea (the 2+1 case, $m_u=m_d$) and in the continuum.

\section{Form factors for $K_{\ell3}$ decay}
\label{sec:form}

The hadronic matrix element between a kaon and a pion
of the weak vector current may be parameterized by two form
factors, $f_+$ and $f_-$, defined by
\be\label{eq:formfacpm}
\langle \pi (p_\pi)\vert V^\mu \vert K(p_K)\rangle =
f_+(q^2) \left(p_K^\mu + p_\pi^\mu\right) +f_- (q^2) \left(p_K^\mu - p_\pi^\mu\right)\ , 
\ee
where $q=p_K-p_\pi$ is the momentum transfer, and $V^\mu=\bar s \gamma^\mu u$
is the appropriate flavor-changing vector current.
It is often more convenient to introduce the scalar form factor $f_0$,
defined by
\be\eqn{f0}
f_0(q^2) = f_+(q^2) +  f_-(q^2) \frac{q^2}{m_K^2-m_\pi^2}\ ,
\ee
in terms of which the matrix element is given by
\be\label{eq:formfac}
\langle \pi (p_\pi)\vert V^\mu \vert K(p_K)\rangle =
f_+(q^2) \left[p_K^\mu + p_\pi^\mu - \frac{m_K^2-m_\pi^2}{q^2}
q^\mu\right]+f_0(q^2)\frac{m_K^2-m_\pi^2}{q^2}q^\mu\ .
\ee
This is useful phenomenologically because it is easier 
to disentangle $f_+$ and $f_0$ experimentally since they are less correlated than $f_+$ and $f_{-}$.
In practice, the key quantity to be calculated on the lattice is the absolute normalization of
the form factor $f_+$ at one value of $q^2$, which is usually taken to be $q^2=0$. Experiments provide 
the relative normalization $f_+(q^2)/f_+(0)$, so once $f_+(0)$ is known, the CKM element $|V_{us}|$
may be extracted from the total $K_{\ell3}$ decay width.  The kinematical relation $f_+(0)=f_0(0)$, which 
follows from \eq{f0}, can be helpful in this regard \cite{Bazavov:2012cd}.  

In this paper we focus on a calculation of $f_+$ in \schpt.  Although the main motivation is to aid 
in the lattice determination at $q^2=0$, we compute $f_+$ at arbitrary $q^2$, since this introduces 
few additional complications.   For definiteness, we consider the mode $K^0\to\pi^- \ell^+\nu$. That mode 
(and its charge conjugate) are the best-measured isospin channels experimentally, and other modes are
usually normalized to it.  For the purposes of the calculation here, the particular isospin 
mode considered is irrelevant, since we do not include electromagnetism, and we do a partially quenched
calculation, so that the valence and sea masses may be chosen arbitrarily. 
For convenience for lattice computations, which usually are performed in
the limit of exact isospin in the sea, we also provide results in that limit. In any case, the effect of isospin violation in the sea is 
NNLO --- higher order than we are considering here.

The form factor $f_+(0)$ can be written as a $\chi$PT expansion:
\ba\label{chptexpansion}
f_+(0) = 1 + f_2 + f_4 + f_6 + ... = 1 + f_2 + \Delta f\,,
\ea 
where the $f_i$ contain corrections of $\order(p^{2i})$ in the chiral 
power counting.
The Ademollo-Gatto (AG) theorem \cite{Ademollo:1964sr}, which follows from vector current conservation, 
ensures that $f_+(0) \to 1$ in the $SU(3)$ limit and, furthermore, that the $SU(3)$  
breaking effects are second order in $(m_K^2-m_\pi^2)$. This fixes $f_2$ in the continuum completely 
in terms of experimental quantities. The AG theorem is a statement about the valence quark masses
in the mesons that enter the weak current (the valence $u$ and $s$), so it remains true as a statement 
about the dependence on valence quark masses even in a
partially quenched theory  \cite{Becirevic:2005py}.  It is straightforward to see this using the approach to 
the AG theorem developed in  \rcite{Sirlin:1979if}, and it follows simply from the ``$U$-spin'' subgroup of
(valence) $SU(3)$ symmetry that rotates the valence $u$ and $s$ into each other. 
Note that although \rcite{Becirevic:2005py} takes the spectator valence quark 
(the valence $d$) to be degenerate with the valence $u$, that is not necessary for the theorem to be valid,
and we work below with arbitrary values of the three valence masses.  Furthermore, since the theorem just
depends on a flavor symmetry, it remains valid when staggered discretization effects are included through 
\schpt. We will verify below that the result of our calculation obeys the theorem.  Nevertheless, violations 
of the AG theorem may be introduced at a later stage in a lattice computation.
In particular, \rcite{Bazavov:2012cd} uses the continuum dispersion relation to relate the matrix element
of the scalar density to that of the vector current of interest, and the dispersion relation is of course
violated on the lattice. The corresponding discretization errors in the AG theorem appear to be  very small, 
however \cite{Bazavov:2012cd}.

\section{Basics of Staggered Chiral Perturbation theory}

\label{sec:preliminaries}

Here, we follow the discussion in \rcite{RMP} fairly closely.
The starting point for \schpt\ is the (Euclidean space) Lee-Sharpe Lagrangian
\cite{LEE_SHARPE} generalized to multiple flavors in 
Ref.~\cite{SCHPT}:
\ba
        \cL  =  \frac{f^2}{8} {\rm Tr}(\partial_{\mu}\Sigma
        \partial_{\mu}\Sigma^{\dagger}) -
        \frac{1}{4}\mu f^2 {\rm Tr}(\cM\Sigma+\cM\Sigma^{\dagger})
        + \frac{m_0^2}{24}({\rm Tr}(\Phi))^2 +a^2\cV \ ,
\label{LSChPT}
\ea
where the meson field $\Phi$, $\Sigma\equiv\exp(i\Phi / f)$, and the quark
mass matrix $\cM$ are $4N_f \times 4N_f$ matrices, $f$ is the pion decay constant at 
LO, and $\mu$ is a low energy constant (LEC). The parameter $a$ is the lattice spacing, 
and discretization effects enter first at $\order(a^2)$. 
We will assume that there are 
three light sea-quark flavors ($u$, $d$, and $s$), 
but $N_f$ in general will be larger than three to accommodate valence quarks 
(and either additional ghost quarks or replicas), in order
 to allow for partial quenching \cite{Bernard:1993sv,REPLICA}.

The field $\Sigma$ transforms
under SU($4N_f$)$_L\times$SU($4N_f$)$_R$ as $\Sigma \rightarrow L\Sigma
R^{\dagger}$.
The field $\Phi$ is given by:
\begin{eqnarray}\label{eq:Phi}
        \Phi = \left( \begin{array}{cccc}
                 U  & \pi^+ & K^+ & \cdots \\*
                 \pi^- & D & K^0  & \cdots \\*
                 K^-  & \bar{K^0}  & S  & \cdots \\*
                \vdots & \vdots & \vdots & \ddots
                \end{array} \right),
\end{eqnarray}
where each entry is a $4\times 4$ matrix in taste space, with, for example,
\begin{equation}\label{eq:sigma}
        \pi^+\equiv\sum_{\Xi=1}^{16} \pi^+_\Xi T_\Xi.
\end{equation}
The 16 Hermitian taste generators $T_\Xi$ are
\begin{equation}\label{eq:T_Xi}
        T_\Xi \in \{ \xi_5, i\xi_{\mu5}, i\xi_{\mu\nu} (\mu>\nu), \xi_{\mu}, I\}.
\end{equation}
Here we use the Euclidean gamma matrices $\xi_{\mu}$, with
$\xi_{\mu\nu}\equiv \xi_{\mu}\xi_{\nu}$ ($\mu <\nu$), $\xi_{\mu5}\equiv \xi_{\mu}\xi_5$, 
and $\xi_I \equiv I$ is the $4\times 4$ identity matrix.
The mass matrix has the form
\begin{eqnarray}
        \cM = \left( \begin{array}{cccc}
                m_u I  & 0 &0  & \cdots \\*
                0  & m_d I & 0  & \cdots \\*
                0  & 0  & m_s I  & \cdots\\*
                \vdots & \vdots & \vdots & \ddots \end{array} \right) \ .
\end{eqnarray}
For generality, we will usually take all sea masses nondegenerate (\ie the 1+1+1 case) below. 
Converting the formulae to the 2+1 ($m_u=m_d$) case, which is relevant for current simulations
\cite{RMP,HISQ-Ensembles}, is straightforward. One-loop results for $f_+(q^2)$ in the 2+1 case 
and in the continuum are compiled in Appendix~\ref{ref:appB}.

The quantity  $m_0$ in Eq.~(\ref{LSChPT}) is  the anomaly
contribution to the mass of the singlet-taste and singlet-flavor meson,
the $\eta'\propto \Tr(\Phi)$.  As usual, the
$\eta'$ decouples in the limit $m_0\to\infty$. However, one may
postpone taking the limit and keep the
$\eta'$
 as a dynamical field \cite{Sharpe:2001fh} in order to avoid putting conditions on the
diagonal elements of $\Phi$.
The diagonal fields, $U,D,\dots$, are then simply the $u\bar u$, $d\bar d$, \dots bound states,
which makes it easy
to follow the quark flow  through chiral diagrams \cite{QUARK-FLOW}
by following the
flavor indices.  A quark flow analysis is particularly useful in our
calculations here because it helps keep track of the many different possible contributions
that all correspond to the same diagram at the chiral (meson) level.

The taste-violating potential $\cV$ in Eq.~(\ref{LSChPT}) is given by
\begin{eqnarray}\label{tastevpot}
        -\cV  & = & C_1
         \Tr(\xi^{(N_f)}_5\Sigma\xi^{(N_f)}_5\Sigma^{\dagger})
+\frac{C_3}{2} [ \Tr(\xi^{(N_f)}_{\nu}\Sigma
        \xi^{(N_f)}_{\nu}\Sigma) + h.c.] \nonumber \\*
        & & +\frac{C_4}{2} [ \Tr(\xi^{(N_f)}_{\nu 5}\Sigma
        \xi^{(N_f)}_{5\nu}\Sigma) + h.c.]
+\frac{C_6}{2}\ \Tr(\xi^{(N_f)}_{\mu\nu}\Sigma
        \xi^{(N_f)}_{\nu\mu}\Sigma^{\dagger}) \nonumber \\*
        & & + \frac{C_{2V}}{4}
                [ \Tr(\xi^{(N_f)}_{\nu}\Sigma)
        \Tr(\xi^{(N_f)}_{\nu}\Sigma)  + h.c.]
        +\frac{C_{2A}}{4} [ \Tr(\xi^{(N_f)}_{\nu
         5}\Sigma)\Tr(\xi^{(N_f)}_{5\nu}\Sigma)  + h.c.] \nonumber \\*
        & & +\frac{C_{5V}}{2} [ \Tr(\xi^{(N_f)}_{\nu}\Sigma)
        \Tr(\xi^{(N_f)}_{\nu}\Sigma^{\dagger})]
         +\frac{C_{5A}}{2} [ \Tr(\xi^{(N_f)}_{\nu5}\Sigma)
        \Tr(\xi^{(N_f)}_{5\nu}\Sigma^{\dagger}) ],
\eqn{V}
\end{eqnarray}
with implicit sums over repeated indices.
The $\xi^{(N_f)}_b$
are block-diagonal $4N_f\times 4N_f$ matrices:
\begin{equation}\label{eq:xi-b}
        \xi^{(N_f)}_b = \left(
        \begin{matrix}
        \xi_b & 0 & 0 & \cdots\cr
        0 & \xi_b & 0 & \cdots\cr
        0 & 0 & \xi_b & \cdots\cr
        \vdots & \vdots & \vdots & \ddots 
        \end{matrix}
        \right)\ ,
\end{equation}
with $\xi_b$ the $4\times 4$ objects,
and $b\in\{5,\mu,\mu\nu\ (\mu<\nu),\mu5,I \}$.

The two-trace terms in $\cV$ generate two-point (``hairpin'') vertices
at $\cO(a^2)$ that mix flavor-neutral particles of
vector and axial tastes. In addition, flavor-neutral, singlet-taste 
particles are mixed by the $m_0^2$ term in Eq.~(\ref{LSChPT}), which
results from the anomaly. For taste $\Xi$, we thus have terms 
the Lagrangian of the form $(\delta_\Xi/2) (U_\Xi + D_\Xi + S_\Xi + \cdots)^2$, where\footnote{Note that
for vector and axial tastes we use 
the notation $\delta_{A,V}$, instead of $\delta'_{A,V}$ used in Ref.~\cite{SCHPT}, 
to avoid  cluttering the notation in the mixed-action case.}
\begin{equation}\label{eq:dp_def}
        \delta_\Xi = \
        \begin{cases} 
        a^2 \delta_V \equiv 16 a^2(C_{2V}-C_{5V})/f^2, &\Xi\in\{\xi_\mu\}\ {\rm (vector\ taste);}\cr
        a^2 \delta_A \equiv 16 a^2(C_{2A}-C_{5A})/f^2, &\Xi\in\{\xi_5\xi_\mu\}\ {\rm (axial\ taste);}\cr
        4m_0^2/3, &\Xi=I\ {\rm (singlet\ taste);}\cr 
        0, &{\rm otherwise.}\cr 
        \end{cases}
\end{equation}
These mixings require us to diagonalize the full mass matrix in each of the
three nontrivial taste channels.
We write the neutral propagator for taste $\Xi$ as:
\begin{eqnarray}\label{eq:prop}
        G_\Xi& =& G_{0,\Xi} + \cD^\Xi \ ,
\end{eqnarray}
where $\cD^\Xi$ is the part of the flavor-neutral propagator of taste $\Xi$ that
is disconnected at the quark level (plus all iterations of intermediate sea quark loops).  Explicitly,
the disconnected propagator for a valence meson $X$ (made out of valence quarks
$x$ and $\bar x$) and $Y$ (similarly composed of $y$ and $\bar y$) is \cite{SCHPT}:
\begin{eqnarray} \label{eq:DiscXi}
        \cD^\Xi_{XY}(p)&=&  -a^2\delta_\Xi
        \frac{ (p^2 + m_{U_\Xi }^2)(p^2 + m_{D_\Xi }^2)(p^2 + m_{S_\Xi }^2) }
        {(p^2 + m_{X_\Xi }^2)(p^2 + m_{Y_\Xi }^2)
        (p^2 + m_{\pi^0_\Xi }^2)(p^2+m_{\eta_\Xi }^2)(p^2 + m_{\eta'_\Xi }^2)}\ .
\end{eqnarray}
Here, $m_{\pi^0_\Xi }^2$, $m_{\eta_\Xi }^2$ and $m_{\eta'_\Xi }^2$ are the
eigenvalues of the full mass-squared matrix  of the neutral sea mesons.  
For the singlet-taste propagator, we may simplify the disconnected propagator by
taking $m_0\to\infty$, and using the fact that, after rooting,
 $m_{\eta'_I}\approx m_0$ for large $m_0$. We obtain
\begin{eqnarray} \label{eq:DiscI}
        \cD^I_{XY}(p)&=&  -\frac{4}{3}
        \frac{ (p^2 + m_{U_I}^2)(p^2 + m_{D_I}^2)(p^2 + m_{S_I}^2) }
        {(p^2 + m_{X_I}^2)(p^2 + m_{Y_I}^2)
        (p^2 + m_{\pi^0_I}^2)(p^2+m_{\eta_I}^2)}\ .
\end{eqnarray}
For the disconnected propagators in the vector-taste and axial-taste case, there is no explicit difference in
\eq{DiscXi} between the rooted and unrooted cases, but the sea masses in the denominators are dependent on the
number of sea-quark flavors coming from intermediate loops, and one should use the appropriate masses in each
case \cite{SCHPT}. 
Note that, when $m_u=m_d$,
the factors of $p^2 + m_{\pi^0}^2$ in \eqs{DiscXi}{DiscI} cancel the factors  of $p^2 + m_{D}^2$.

At leading order in \schpt, the mass of a pseudoscalar meson (``pion'') of 
taste $\Xi$ made of quarks with flavor $i,j$ is given by
\ba\label{eq:mesonmass}
M_{ij,\Xi}^2 = \mu(m_i+m_j) + a^2\Delta_\Xi\,,
\ea
where $\mu$ is the LEC in Eq.~(\ref{LSChPT}), and 
$\Delta_\Xi$ is the taste splitting, which can be written as a linear combination of 
the LECs $C_1$, $C_3$, $C_4$, and $C_6$ in Eq.~(\ref{tastevpot}). The splitting vanishes 
for the pseudoscalar-taste pion ($\Delta_5=0$), so it is a true Goldstone boson
of the lattice theory in the chiral limit. The standard staggered power counting, which 
we follow here, assumes that the taste splittings and squared Goldstone
pion masses are  comparable.  This is true of not only for the MILC asqtad ensembles, 
but also for the MILC HISQ ensembles, which have
smaller taste splittings but have smaller quark masses as well. 
Schematically, one describes the power counting by saying
$a^2 \sim m$, where $m$ is a generic quark mass.

\section{Staggered chiral perturbation theory with a mixed staggered action }

\label{sec:mixed}

We now turn to the case of a mixed staggered theory, where the actions for the sea quarks and for
the valence quarks are different, although both are staggered.  We work out the staggered chiral 
Lagrangian for this case by starting with the quark-level Symanzik effective theory, which
encodes the discretization errors as (higher dimensional) continuum operators.   Once we have the
Symanzik theory, it is straightforward to find the corresponding chiral theory, following a  
``spurion" analysis.  This is the standard approach for including discretization errors in a chiral theory,
first introduced by Sharpe and collaborators in Refs.~\cite{Sharpe:1998xm,LEE_SHARPE}.
We note that, after we worked out the properties of
staggered mixed-action chiral perturbation theory, we discovered Ref.~\cite{Bae:2010ki}, which 
developed mixed  action staggered chiral perturbation theory several years ago, and found
many of the results that are given in this section.

\subsection{Symanzik effective theory}\label{sec:SET}
Both the asqtad and the HISQ quarks have the full staggered set of symmetries.  In particular
they have separate $U(1)_\epsilon$ for each flavor, and there are overall rotation and shift
symmetries.  Rotations and shifts must be done on all staggered fields at once, since the 
the gluons must also be transformed.  The analysis of the $\cO(a^2)$ taste-violating four-quark
operators in the Symanzik effective theory (SET) is then completely standard, and
closely parallels the discussion in \rcite{RMP}.  Recall that, in an ordinary (unmixed)
staggered theory with $n$ flavors, the four-quark operators are of the form
\begin{equation}\eqn{SET-normal}
 a^2  \bar q_i(\gamma_s \otimes \xi_t) q_i\;  \bar q_j(\gamma_s \otimes \xi_t) q_j \ ,
\end{equation}
where $i,j$ are (summed) flavor indices,
and by $U(1)_\epsilon$  symmetry, the spin $\otimes$ taste
combination $\gamma_s \otimes \xi_t$ must be odd with respect to $\gamma_5 \otimes \xi_5$.
An example is
tensor (T) $\otimes$ vector (V):  $\gamma_{\mu\nu} \otimes \xi_\lambda$.  In 
``type A'' operators, the indices on the two $\gamma_s$ matrices in \eq{SET-normal}
are the same, as
are the indices on the two $\xi_t$ matrices, but there are no indices in 
common between spin and
taste.  In ``type B'' operators some indices are repeated four times and are common to 
both spin and taste matrices.  Type B operators break Euclidean invariance. They turn 
out to be irrelevant for the
LO chiral theory because their chiral representatives require extra derivatives; they
first appear at
NLO \cite{LEE_SHARPE}.

In the mixed asqtad-HISQ theory, the operators are basically the same as above,
but they come in three varieties: valence-valence, sea-sea, and sea-valence. Defining
$P_v$ and $P_\sigma$ as projectors on the valence and sea quarks, we have, instead of
\eq{SET-normal}:
\begin{eqnarray}
&c_{vv}\, a^2\,  \bar q(\gamma_s \otimes \xi_t)P_v q\;  \bar q(\gamma_s \otimes \xi_t) P_vq 
\nonumber \\
+&c_{\sigma\sigma}\, a^2\,  \bar q(\gamma_s \otimes \xi_t)P_\sigma q\;  \bar q(\gamma_s \otimes \xi_t) P_\sigma q \eqn{SET-mixed}\\
+&2c_{v\sigma}\, a^2\,  \bar q(\gamma_s \otimes \xi_t)P_v q\;  \bar q(\gamma_s \otimes \xi_t) P_\sigma q 
\nonumber \ ,
\end{eqnarray}
where flavor indices are now implied.  The key point is that there are independent
coefficients $c_{vv}$,
$c_{\sigma\sigma}$, and $c_{v\sigma}$, because there
is no lattice symmetry that turns valence and sea quarks into each other.  The normalization
in \eq{SET-mixed} is chosen so that the ``unmixed limit'' in which the valence and sea actions
are identical is $c_{vv}=c_{\sigma\sigma}=c_{v\sigma}$.  This follows from \eq{SET-normal},
if the flavor indices $i,j$ are simply allowed to run over all quarks, both valence and sea.

\subsection{Mixed theory chiral Lagrangian at leading order}\label{sec:lagrangian}

The meson field $\phi$ is a matrix in flavor-taste space, where the flavor indices 
run over valence and sea indices.\footnote{We will assume that we will use the quark 
flow or the replica method \cite{REPLICA} to remove the valence-quark determinant; 
any needed replica indices in the latter case will be implicit.} 
We write
\begin{equation}\label{eq:phi}
        \quad \phi\equiv\sum_{a=1}^{16} \phi^a T_a   \ ,
\end{equation}
where $\phi^a$ is itself a matrix in flavor (really flavor-replica) space, and
the Hermitian taste generators $T_a$ are the same as in Eq.~(\ref{eq:T_Xi}).


The chiral matrix $\Sigma$ is defined as is in Eq.~(\ref{LSChPT}); it transforms
under $SU(4n)_L\times SU(4n)_R$ as $\Sigma \rightarrow L\Sigma
R^{\dagger}$, where $n$ here denotes the total number of flavors (flavor-replicas) of 
all types. Valence quarks will be denoted by $x$, $y$, \dots, and corresponding neutral 
valence mesons will be $X$, $Y$, \dots, where the taste has not been 
specified.  Similarly, sea quarks will be denoted $u$, $d$, $s$ \dots, and 
corresponding neutral valence mesons will be $U$, $D$, $S$ \dots.

The spurion analysis to derive the chiral Lagrangian parallels the normal staggered case. 
Each taste spurion from \eq{SET-mixed} now comes with  an additional projector (either $P_v$ or $P_\sigma$)
which leads to a 3-fold increase in the taste-violating
chiral terms: They will be either $vv$, $\sigma\sigma$,
or $v\sigma$, and their  coefficients will be independent. To find these terms, 
we just have to take the normal taste-violating potential $\cV$ 
\cite{SCHPT} and insert appropriate projectors~\footnote{An exception occurs 
in the  special case of singlet-taste operators, treated below.}. 
We will denote the corresponding taste-violating potentials as $\cV_{vv}$, 
$\cV_{\sigma\sigma}$ and $\cV_{v\sigma}$.
These potentials are given by:
\begin{eqnarray}
\cV_{vv} &=& \cU_{vv}+\cU'_{vv}\;; \nonumber \\
        -\cU_{vv}  & = & C^{vv}_1\;
         {\rm Tr}(\xi_5P_v\Sigma\xi_5P_v\Sigma^{\dagger}) 
+\frac{C^{vv}_3}{2}\; [ {\rm Tr}(\xi_{\mu}P_v\Sigma
        \xi_{\mu}P_v\Sigma) + p.c.] \nonumber \\*
        & & +\frac{C^{vv}_4}{2}\; [ {\rm Tr}(\xi_{\mu 5}P_v\Sigma
        \xi_{5\mu}P_v\Sigma) + p.c.] 
+\frac{C^{vv}_6}{2}\; {\rm Tr}(\xi_{\lambda\mu}P_v\Sigma
        \xi_{\mu\lambda}P_v\Sigma^{\dagger}) \nonumber \\*
         -\cU'_{vv}  & = & \frac{C^{vv}_{2V}}{4} \;
                [ {\rm Tr}(\xi_{\mu}P_v\Sigma) {\rm Tr}(\xi_{\mu}P_v\Sigma)  + p.c.] 
        +\frac{C^{vv}_{2A}}{4}\; [ {\rm Tr}(\xi_{\mu
         5}P_v\Sigma){\rm Tr}(\xi_{5\mu}P_v\Sigma)  + p.c.] \nonumber \\*
        & & +\frac{C^{vv}_{5V}}{2}\; [ {\rm Tr}(\xi_{\mu}P_v\Sigma)
        {\rm Tr}(\xi_{\mu}P_v\Sigma^{\dagger})]
         +\frac{C^{vv}_{5A}}{2}\; [ {\rm Tr}(\xi_{\mu5}P_v\Sigma)
        {\rm Tr}(\xi_{5\mu}P_v\Sigma^{\dagger}) ]\ , 
\eqn{Vvv}
\end{eqnarray}
\begin{eqnarray}
\cV_{\sigma\sigma} &=& \cU_{\sigma\sigma}+\cU'_{\sigma\sigma}\;; \nonumber \\
        -\cU_{\sigma\sigma}  & = & C^{\sigma\sigma}_1\;
         {\rm Tr}(\xi_5P_\sigma\Sigma\xi_5P_\sigma\Sigma^{\dagger}) 
+\frac{C^{\sigma\sigma}_3}{2}\; [ {\rm Tr}(\xi_{\mu}P_\sigma\Sigma
        \xi_{\mu}P_\sigma\Sigma) + p.c.] \nonumber \\*
        & & +\frac{C^{\sigma\sigma}_4}{2}\; [ {\rm Tr}(\xi_{\mu 5}P_\sigma\Sigma
        \xi_{5\mu}P_\sigma\Sigma) + p.c.] 
+\frac{C^{\sigma\sigma}_6}{2}\; {\rm Tr}(\xi_{\lambda\mu}P_\sigma\Sigma
        \xi_{\mu\lambda}P_\sigma\Sigma^{\dagger}) \nonumber \\*
         -\cU'_{\sigma\sigma}  & = & \frac{C^{\sigma\sigma}_{2V}}{4} \;
                [ {\rm Tr}(\xi_{\mu}P_\sigma\Sigma) {\rm Tr}(\xi_{\mu}P_\sigma\Sigma)  + p.c.] 
        +\frac{C^{\sigma\sigma}_{2A}}{4}\; [ {\rm Tr}(\xi_{\mu
         5}P_\sigma\Sigma){\rm Tr}(\xi_{5\mu}P_\sigma\Sigma)  + p.c.] \nonumber \\*
        & & +\frac{C^{\sigma\sigma}_{5V}}{2}\; [ {\rm Tr}(\xi_{\mu}P_\sigma\Sigma)
        {\rm Tr}(\xi_{\mu}P_\sigma\Sigma^{\dagger})]
         +\frac{C^{\sigma\sigma}_{5A}}{2}\; [ {\rm Tr}(\xi_{\mu5}P_\sigma\Sigma)
        {\rm Tr}(\xi_{5\mu}P_\sigma\Sigma^{\dagger}) ]\ , 
\eqn{Vss}
\end{eqnarray}
\begin{eqnarray}
\cV_{v\sigma}&=& \cU_{v\sigma}+\cU'_{v\sigma}\;; \nonumber \\
        -\cU_{v\sigma}  & = & C^{v\sigma}_1\;
         [{\rm Tr}(\xi_5P_v\Sigma\xi_5P_\sigma\Sigma^{\dagger}) + p.c.]
+C^{v\sigma}_3\; [ {\rm Tr}(\xi_{\mu}P_v\Sigma
        \xi_{\mu}P_\sigma\Sigma) + p.c.] \nonumber \\*
        & & +C^{v\sigma}_4\; [ {\rm Tr}(\xi_{\mu 5}P_v\Sigma
        \xi_{5\mu}P_\sigma\Sigma) + p.c.] 
+\frac{C^{v\sigma}_6}{2}\; [ {\rm Tr}(\xi_{\lambda\mu}P_v\Sigma
        \xi_{\mu\lambda}P_\sigma\Sigma^{\dagger})+ p.c.] \nonumber \\*
         -\cU'_{v\sigma}  & = & \frac{C^{v\sigma}_{2V}}{2} \;
                [ {\rm Tr}(\xi_{\mu}P_v\Sigma) {\rm Tr}(\xi_{\mu}P_\sigma\Sigma)  + p.c.] 
        +\frac{C^{v\sigma}_{2A}}{2}\; [ {\rm Tr}(\xi_{\mu
         5}P_v\Sigma){\rm Tr}(\xi_{5\mu}P_\sigma\Sigma)  + p.c.] \nonumber \\*
        & &\hspace{-1cm} +\frac{C^{v\sigma}_{5V}}{2}\; [ {\rm Tr}(\xi_{\mu}P_v\Sigma)
        {\rm Tr}(\xi_{\mu}P_\sigma\Sigma^{\dagger})+p.c.]
         +\frac{C^{v\sigma}_{5A}}{2}\; [ {\rm Tr}(\xi_{\mu5}P_v\Sigma)
        {\rm Tr}(\xi_{5\mu}P_\sigma\Sigma^{\dagger})+p.c. ]\ .
\eqn{Vvs}
\end{eqnarray}
As usual \cite{SCHPT}, we denote the 
one-trace terms that contribute to tree-level mass splittings by $\cU$, and the 
two-trace terms that give rise to the flavor-singlet taste-violating hairpins by $\cU'$.  
The notation ``p.c.''\ implies the parity conjugate ($\Sigma \leftrightarrow \Sigma^\dagger$).
It sometimes gives a different result from Hermitian conjugation (\eg in the 
$C_1^{v\sigma}$ terms, where it has the effect of switching
$P_v \leftrightarrow P_\sigma$); the potentials are of course still
Hermitian.  Note that the size of each operator in $\cU'_{v\sigma}$ has been ``doubled''
relative to the corresponding operators in  $\cU'_{vv}$ or  $\cU'_{\sigma\sigma}$; this
is accomplished either by changing the overall coefficient or by adding the parity 
conjugate where it is needed.  The normalization is convenient because then the unmixed limit  
where sea and valence actions are the same becomes $C_k^{vv}=C_k^{\sigma\sigma}=C_k^{v\sigma}$, 
where $k\in\{1,3,4,6,2V,2A,5V,5A\}$.\footnote{This follows from the comments on normalization 
following \eq{SET-mixed}.}

The case where $\xi_t=I$ in \eq{SET-mixed}, 
\ie singlet-taste bilinears, is special.  These operators don't contribute in a normal (unmixed) staggered theory 
because the taste spurion in \eq{SET-normal}
is the identity, so they only generate trivial (constant) chiral
operators. However, for the mixed theory, the spurion is a projector ($P_v$ or $P_\sigma$). Therefore these operators will generate
chiral operators that have no counterpart in the unmixed case. 
Since $\xi_t=I$, the spin $\gamma_s$ in
 \eq{SET-mixed} is V or A.  Thus, the chiral operators will be similar to those generated by V$\times$P or A$\times$P. They
give a chiral operator proportional to $\Tr(\xi_5\Sigma\xi_5\Sigma^{\dagger})$ in the unmixed case, so here we just need to replace
each $\xi_5$ by either $P_v$ or $P_\sigma$.  The result is
\ba\eqn{C0term}
 C^{vv}_0\; {\rm Tr}(P_v\Sigma P_v\Sigma^{\dagger}) +  C^{\sigma\sigma}_0\; {\rm Tr}(P_\sigma\Sigma P_\sigma\Sigma^{\dagger})+
 C^{v\sigma}_0\; [{\rm Tr}(P_v\Sigma P_\sigma\Sigma^{\dagger})+ {\rm Tr}(P_\sigma\Sigma P_v\Sigma^{\dagger})] \ ,
\ea
where the form of the $ C^{v\sigma}_0$ term is required by parity invariance. 

The operator in \eq{C0term} can be simplified because various linear
combinations of the terms are just constants. Defining 
$P_\pm\equiv P_\sigma\pm P_v$, and inserting $P_\sigma=\half(P_++P_-)$,
$P_v=\half(P_+-P_-)$ in \eq{C0term}, we see that any term involving $P_+$ reduces
to a constant (independent of the chiral fields), 
because $P_+$ is the identity, and $\Sigma\Sigma^\dagger=I$.
With the more standard notation $P_-\equiv\tau_3$ \cite{hep-lat/0306021,hep-lat/0503009},
this operator becomes
\begin{equation}
\eqn{Cmix}
-C_{\rm mix}\Tr(\tau_3\Sigma \tau_3\Sigma^{\dagger}) \,
\end{equation}
where $C_{\rm mix} \equiv (2 C^{v\sigma}_0-  C^{\sigma\sigma}_0 -C^{vv}_0 )/4$ is a measure
of the mismatch between the sea and valence actions, and would vanish if there were a
lattice symmetry interchanging sea and valence quarks.


The leading-order Euclidean Lagrangian, in analogy with the unmixed case given 
in Eq.~(\ref{LSChPT}), is then
\begin{eqnarray}
	\cL_{LO} &=& \frac{f^2}{8} {\rm Tr}(\partial_{\mu}\Sigma
	\partial_{\mu}\Sigma^{\dagger}) - \frac{1}{4}B\, f^2\,
	{\rm Tr}(\cM\Sigma + \cM\Sigma^{\dagger})
+ \frac{2m_0^2}{3}\,\phi_I^2 \nonumber \\
&&-a^2C_{\rm mix}{\rm Tr}(\tau_3\Sigma \tau_3\Sigma^{\dagger}) + a^2\cV_{vv}+a^2\cV_{\sigma\sigma}+a^2\cV_{v\sigma},\label{eq:LLO} 
\end{eqnarray}
where, again, $\cM$ is the quark mass matrix, $\mu$ is a LEC that relates quark 
and meson masses, and $m_0$ is the $\eta'$ mass term from the anomaly. The $\eta'$ field 
in Eq.~(\ref{eq:LLO}) is defined by $\phi_I = X_I+ Y_I + \cdots + U_I+ D_I + S_I \cdots$, where 
the subscript $I$ indicates the taste singlet.

As usual the potentials have ``accidental'' $SO(4)$ taste symmetry \cite{LEE_SHARPE},
so all mesons fall into $SO(4)$ representations with tastes $P$, $A$, $T$, $V$, and $I$.
The valence-valence and sea-sea mesons with non-singlet flavor get standard mass splittings
from $\cU_{vv}$ and $\cU_{\sigma\sigma}$, respectively.  The splittings of the mixed
(valence-sea)
mesons come not only from  the $C_{\rm mix}$ term and $\cU_{v\sigma}$, however, but also from 
$\cU_{vv}$ and $\cU_{\sigma\sigma}$ \cite{arXiv:0905.2566}. Expanding to quadratic order,
we find that a mixed meson with valence flavor $q$, sea flavor $\mathscr{S}$, and taste $\Xi$ 
gets mass
\begin{equation}\label{eq:tree-mass}
        M_{q\mathscr{S},\Xi}^2 = \mu(m_q+m_\mathscr{S}) 
                +a^2\Delta^{v\sigma}_\Xi,
\end{equation} 
where the splittings $\Delta^{v\sigma}_\Xi$ are given by:
\begin{eqnarray}\label{eq:split-vs}
        \Delta^{v\sigma} (\xi_5) & \equiv & \Delta^{v\sigma}_P = \frac{4}{f^2}\Big[
4C_{\rm mix}-2C_1^{v\sigma}+C_1^{vv}+C_1^{\sigma\sigma} 
-8C_3^{v\sigma}+4C_3^{vv}+4C_3^{\sigma\sigma} \nonumber \\*
&&\hspace{2cm} -8C_4^{v\sigma}+4C_4^{vv}+4C_4^{\sigma\sigma}
-12C_6^{v\sigma}+6C_6^{vv}+6C_6^{\sigma\sigma}
\Big]
                \nonumber \\*
        \Delta^{v\sigma} (\xi_{\mu5}) & \equiv & \Delta^{v\sigma}_A = \frac{4}{f^2}\Big[ 
4C_{\rm mix}+2C_1^{v\sigma}+C_1^{vv}+C_1^{\sigma\sigma}
+4C_3^{v\sigma}+4C_3^{vv}+4C_3^{\sigma\sigma}\nonumber \\*
&&\hspace{2cm} -4C_4^{v\sigma}+4C_4^{vv}+4C_4^{\sigma\sigma}+
6C_6^{vv}+6C_6^{\sigma\sigma}
        \Big] \nonumber \\*
        \Delta^{v\sigma} (\xi_{\mu\nu})  & \equiv &\Delta^{v\sigma}_T =
                \frac{4}{f^2}\Big[4C_{\rm mix}-2C_1^{v\sigma}+C_1^{vv}+C_1^{\sigma\sigma}
+4C_3^{vv}+4C_3^{\sigma\sigma}\nonumber \\*
&&\hspace{2cm} +4C_4^{vv}+4C_4^{\sigma\sigma}
+4C_6^{v\sigma}+6C_6^{vv}+6C_6^{\sigma\sigma}
\Big] \nonumber \\*
        \Delta^{v\sigma} (\xi_{\mu}) & \equiv & \Delta^{v\sigma}_V = \frac{4}{f^2}\Big[
4C_{\rm mix}+2C_1^{v\sigma}+C_1^{vv}+C_1^{\sigma\sigma}
-4C_3^{v\sigma}+4C_3^{vv}+4C_3^{\sigma\sigma}\nonumber \\*
&&\hspace{2cm} +4C_4^{v\sigma}+4C_4^{vv}+4C_4^{\sigma\sigma}
+6C_6^{vv}+6C_6^{\sigma\sigma}
\Big] \nonumber \\*
        \Delta^{v\sigma} (\xi_I)  & \equiv & \Delta^{v\sigma}_I =
         = \frac{4}{f^2}\Big[
4C_{\rm mix}-2C_1^{v\sigma}+C_1^{vv}+C_1^{\sigma\sigma}
+8C_3^{v\sigma}+4C_3^{vv}+4C_3^{\sigma\sigma}\nonumber \\*
&&\hspace{2cm} +8C_4^{v\sigma}+4C_4^{vv}+4C_4^{\sigma\sigma}
-12C_6^{v\sigma}+6C_6^{vv}+6C_6^{\sigma\sigma}
\Big].
\end{eqnarray}
In contrast, the splittings for a valence-valence meson are given by
\begin{eqnarray}\label{eq:split-vv}
        \Delta^{vv} (\xi_5) & \equiv & \Delta^{vv}_P = 0
                \nonumber \\*
        \Delta^{vv} (\xi_{\mu5}) & \equiv & \Delta^{vv}_A = \frac{16}{f^2}\left( 
        C^{vv}_1 + 3C^{vv}_3 + C^{vv}_4 + 3C^{vv}_6 \right) \nonumber \\*
        \Delta^{vv} (\xi_{\mu\nu})  & \equiv &\Delta^{vv}_T =
                \frac{16}{f^2}\left(2C^{vv}_3 + 2C^{vv}_4 + 4C^{vv}_6\right) \nonumber \\*
        \Delta^{vv} (\xi_{\mu}) & \equiv & \Delta^{vv}_V = \frac{16}{f^2}\left( 
        C^{vv}_1 + C^{vv}_3 + 3C^{vv}_4 + 3C^{vv}_6 \right) \nonumber \\*
        \Delta^{vv} (\xi_I)  & \equiv & \Delta^{vv}_I 
         = \frac{16}{f^2}\left( 
        4C^{vv}_3 + 4C^{vv}_4 \right).
\end{eqnarray}
Sea-sea mesons obey equations  identical to \eq{split-vv} but
with $vv\to\sigma\sigma$ everywhere.

Note that
in general a pseudoscalar-taste mixed meson is not a Goldstone boson because the required
axial symmetry would interchange valence and sea quarks and is not a lattice
symmetry in the mixed action case. Thus its mass has a non-zero contribution 
proportional to $a^2$ and independent of the quark masses,
unlike the pseudoscalar-taste valence-valence or sea-sea mesons. 
 For the mixed P meson, this contribution would
vanish if $C_{\rm mix}=0$ and $C_k^{v\sigma} =C_k^{vv}=C_k^{\sigma\sigma}$ for
$k\in\{1,3,4,6,2V,2A,5V,5A\}$;
this would be required if sea and valence had the same action, but not
otherwise.  One can also check from \eqs{split-vs}{split-vv}
that under these conditions all splittings for valence-valence, sea-sea, and valence-sea
mesons are identical.

Upon expanding $\cU_{vv}'$,  $\cU_{v\sigma}'$ and  $\cU_{\sigma\sigma}'$ in 
\eqsthree{Vvv}{Vss}{Vvs} to quadratic order, we find two-point vertices (``hairpins'')
coupling
flavor-neutral mesons for both axial tastes and vector tastes.  In particular, in the
axial case, there are the following quadratic terms in $\cL_{LO}$:
\begin{eqnarray}
&\frac{1}{2}a^2\delta^{vv}_A\;(X_{\mu5}+Y_{\mu5}+\cdots)^2\;;	
&\hspace{5mm} \delta^{vv}_A \equiv \frac{16}{f^2}(C_{2A}^{vv}-C_{5A}^{vv})\ , \eqn{dAvv}\\
&\frac{1}{2}a^2\delta^{\sigma\sigma}_A\;(U_{\mu5}+D_{\mu5}+S_{\mu5}+\cdots)^2\;;	
&\hspace{5mm} \delta^{\sigma\sigma}_A \equiv \frac{16}{f^2}(C_{2A}^{\sigma\sigma}-C_{5A}^{\sigma\sigma})\;,\eqn{dAss} \\
&a^2\delta^{v\sigma}_A\;(X_{\mu5}+Y_{\mu5}+\cdots)(U_{\mu5}+D_{\mu5}+S_{\mu5}+\cdots)\;;	
&\hspace{5mm} \delta^{v\sigma}_A \equiv \frac{16}{f^2}(C_{2A}^{v\sigma}-C_{5A}^{v\sigma}) \ . \eqn{dAvs}
\end{eqnarray}
Taking into account 
the minus sign from $e^{-S}$, this means that the two-point vertex coupling
$X_{\mu5}$ and $Y_{\mu5}$ is $-\delta^{vv}_A$; that coupling 
$U_{\mu5}$ and $D_{\mu5}$ is $-\delta^{\sigma\sigma}_A$; and that coupling
$X_{\mu5}$ and $U_{\mu5}$ is $-\delta^{v\sigma}_A$, \etc  The vector-taste case is similar,
with simply $A\to V$ and $\mu_5\to\mu$.

There are also standard hairpins in the singlet-taste channel, coming from the anomaly
($m_0^2$) term in $\cL_{LO}$. This produces a vertex of strength $-4m_0^2/3$ between all
singlet-taste, flavor-neutral mesons,
\eg between $X_I$ and $Y_I$, 
between $U_{I}$ and $D_{I}$, and between $X_I$ and
$U_I$. Note that, unlike the unphysical taste-violating hairpins above,
 there is no possibility of different strengths
in the sea and valence sectors
for this physical singlet-taste vertex, even after
including $\cO(a^2)$ corrections. The reason is that the (non-anomalous)
continuum chiral symmetries
require that the anomaly term can only be a function of the equally weighted sum
$X_{I}+Y_{I}+\cdots+U_{I}+D_{I}+S_{I}+\cdots \propto \tr\ln(\Sigma)=
\ln\det(\Sigma)$, not $X_{I}+Y_{I}+\cdots$ or $U_{I}+D_{I}+S_{I}+\cdots$
separately.  At $\cO(a^2)$, any new operators must be constructed from $\Sigma$,
$\Sigma^\dagger$, and the spurions.  Separate functions of  $X_{I}+Y_{I}+\cdots$ or 
$U_{I}+D_{I}+S_{\mu5}+\cdots$ are again forbidden, basically because
the determinant of a projector vanishes.\footnote{We thank M.\ Golterman for discussions
on this issue.}

The disconnected propagators in various taste channels can then be found by summing
the geometric series with sea quark loop insertions, as usual.  In the singlet-taste 
channel, this is completely standard, since the hairpin vertex is universal.
The result for a disconnected propagator between valence meson $X_I$ and 
valence meson $Y_I$ has the same form as in the unmixed case, Eq.~(\ref{eq:DiscI}). 


The only effect of the mixed action in Eq.~(\ref{eq:DiscI}) is that the 
splittings that contribute
to the valence meson masses $m_{X_I}$ and $m_{Y_I}$ are different from the splittings
contributing to the sea mesons $m_{U_I}$,  $m_{\pi_I}$, \etc  In fitting to chiral forms
that result from mixed S\chpt\, one should measure these splittings and input them:
In the application in Ref.~\cite{Bazavov:2012cd}, 
they are just the valence HISQ splittings, and the normal asqtad sea splittings.

The axial-taste and vector-taste cases are more complicated because of the 
presence of three different hairpin coefficients in each channel, \eqsthree{dAvv}{dAss}{dAvs}.
In the series defining the disconnected vector-taste propagator, $D^V_{XY}(p^2)$, the first
term, with no sea quark loops, is proportional to $\delta_V^{vv}$ (times two
valence propagators) since the valence
mesons couple to each other directly.  The next term
is proportional to $(\delta_V^{v\sigma})^2$, since the valence
mesons couple to the single sea quark loop.  After that, each additional sea quark loop brings
in a factor of  $\delta_V^{\sigma\sigma}$, as the sea mesons couple to each other,
in addition to the overall  $(\delta_V^{v\sigma})^2$ factor.  Thus, all terms except
the first form a geometric series that, aside from an overall factor of
  $(\delta_V^{v\sigma}/ \delta_V^{\sigma\sigma})^2$, is identical to the corresponding
series in the pure sea (asqtad) theory.  There is a mismatch in the first term, however,
which would be proportional to
$(\delta_V^{v\sigma})^2/ \delta_V^{\sigma\sigma}$ if the same correspondence held.  Thus
we need to add and subtract a term like the first one, but with a factor of
$(\delta_V^{v\sigma})^2/ \delta_V^{\sigma\sigma}$ instead of  $\delta_V^{vv}$.
The result is:

\begin{eqnarray}
	\cD^V_{XY}(p)&=& 
-\frac{ a^2 (\delta_V^{v\sigma})^2/ \delta_V^{\sigma\sigma}}{(p^2 + m_{X_V}^2)(p^2 + m_{Y_V}^2)}\ 
\frac{(p^2 + m_{U_V}^2)(p^2 + m_{D_V}^2)(p^2 + m_{S_V}^2)}
		{(p^2 + m_{\pi_V}^2)(p^2 + m_{\eta_V}^2)(p^2 + m_{\eta'_V}^2)} \nonumber \\
&&\quad- \frac{a^2[\delta_V^{vv} -(\delta_V^{v\sigma})^2/ \delta_V^{\sigma\sigma}]}{(p^2 + m_{X_V}^2)(p^2 + m_{Y_V}^2)}\ .  \eqn{DA} 
\end{eqnarray}
For the axial-taste channel, just let $V\to A$ everywhere.

Thus there is a ``normal'' hairpin term with strength $ (\delta_\Xi^{v\sigma})^2/ \delta_\Xi^{\sigma\sigma}$ (with $\Xi=A,V$), plus an additional product of two poles proportional
to $\delta_\Xi^{vv} -(\delta_\Xi^{v\sigma})^2/ \delta_\Xi^{\sigma\sigma}$. Note that
this additional term means that there is a real double pole (for $m_{X_\Xi}=m_{Y_\Xi}$) 
even if the quark masses are tuned to the limit where valence-valence
and sea-sea masses are equal in the taste-$\Xi$ (or any other taste) channel.\footnote{In practice,
 it is most common to tune valence-valence
and sea-sea masses equal in the pseudoscalar-taste (Goldstone) channel.}
This is an example of the sickness
of a mixed-action theory; of course it goes away in the continuum limit.

A ``factorization'' assumption about the four-quark operators
suggests a natural size for the mixed hairpin coefficients $\delta_\Xi^{v\sigma}$.
Factorization is the assumption that the four quark operators take the form of squares
of bilinears: $(c_v \bar q P_v q + c_\sigma \bar q P_\sigma q)^2$. 
One would then expect $\delta_\Xi^{v\sigma}\sim
\sqrt{\delta_\Xi^{vv}\delta_\Xi^{\sigma\sigma}}$. 
Note that this only an order of magnitude argument; even if
factorization were exact for every four-quark operator, $\delta_\Xi^{v\sigma}$ would
not equal $\sqrt{\delta_\Xi^{vv}\delta_\Xi^{\sigma\sigma}}$ unless
the ratio $c_\sigma/c_v$ were also identical for each
operator contributing to the hairpins.  With this guess for the size of the $\delta_\Xi$
parameters, we expect that the coefficient of the normal disconnected propagator 
is $\sim \delta_\Xi^{vv}$, 
\ie comparable to a typical HISQ taste-splitting.  This estimate also
suggests that the coefficient of the new double-pole term may
be rather small.  It is therefore convenient to
define the parameter $\delta_\Xi^{{\rm mix}} =
\delta^{vv}_\Xi-(\delta^{v\sigma}_{\Xi})^2/\delta^{\sigma\sigma}_\Xi$ and write the
S$\chi$PT expressions in terms of $\delta_\Xi^{{\rm mix}}$ (which we expect to be
suppressed with respect to the parameters in the sea and the valence sectors)
and $\delta^{vv}_\Xi$:
\begin{eqnarray}\eqn{mixed-DiscXi}
{\cal D}^{\Xi}_{XY}(p) & = & -\frac{a^2(\delta^{vv}_\Xi-\delta^{{\rm mix}}_\Xi)}
{(p^2+m_{X_\Xi}^2)(p^2+m_{Y_\Xi}^2)}
\frac{(p^2+m_{U_\Xi}^2)(p^2+m_{D_\Xi}^2)(p^2+m_{S_\Xi}^2)}
{(p^2+m_{\pi_\Xi}^2)(p^2+m_{\eta_\Xi}^2)(p^2+m_{\eta'_\Xi}^2)}\nonumber\\
&& -\frac{a^2\delta^{{\rm mix}}_\Xi}{(p^2+m_{X_I}^2)(p^2+m_{Y_I}^2)}\,\,.
\end{eqnarray}
We emphasize that \eq{mixed-DiscXi} is for the vector-taste and axial-taste cases only; the
mixed-action singlet-taste disconnected propagator has the identical form to the unmixed
version, either \eq{DiscXi} for $\Xi=I$, or \eq{DiscI} after rooting and $m_0\to\infty$. 
Although available data is not enough to determine the value of 
$\delta_\Xi^{mix}$ precisely, current chiral fits to data using the asqtad action in 
the sea sector and the HISQ action in the valence sector~\cite{Bazavov:2012cd} 
prefer non-zero values that 
are of the same sign but an order of magnitude smaller than $\delta_\Xi^{vv}$.

The partially quenched one-loop calculation involves mesons made of two valence, two
sea, and one valence and one sea quarks. The corresponding masses are given by the
definition in Eq.~(\ref{eq:mesonmass}) with the taste splittings $\Delta_\Xi$ being
$\Delta_\Xi^{vv}$, $\Delta_\Xi^{\sigma\sigma}$, and $\Delta_\Xi^{v\sigma}$, respectively.

We note that  the taste-violating neutral propagator in the staggered mixed-action
theory,  \eq{DA},
has been derived previously by Bae \et\ \cite{Bae:2010ki}, as have 
our estimates for the natural size of the new mixed-action hairpin parameters.

\section{One-loop Calculation of $f_+(q^2)$}

\label{sec:one-loop}

Here we perform the chiral calculation of the form factor $f_+(q^2)$ at one-loop  
in standard \schpt, that is, with an unmixed staggered action.  In \secref{mixed-results}, 
we give the results again (but not the detailed calculation) for the mixed-action case. 
The NLO contribution for $q^2\ne0$ is given by the sum of the one-loop contributions 
that we denote $f_2(q^2)$ in analogy to Eq.~(\ref{chptexpansion}), plus an 
analytical term $\frac{4}{f^2}L_9^r(\mu) q^2$, where $L_9^r(\mu)$ is a renormalized NLO  
low energy constant and $\mu$ the scale at which the logarithms in $f_2(q^2)$ are evaluated. 
In the rest of the paper we omit this contribution in the analytical expressions but 
it should be added to our results for $f_2(q^2)$ for applications. 
We focus on the decay $K^0\to\pi^-\ell^+\nu$, which, at the quark level, is due to the charged
weak vector current $\bar s\gamma_\mu u$.  Allowing for partial quenching,
we let $\bar y$  be the valence antiquark corresponding to $\bar s$,  $\bar x$ be the valence
antiquark corresponding to $\bar u$, and $x'$ be the spectator quark corresponding to $d$. 
Thus the decaying valence pseudoscalar is $x'\bar y$, the outgoing pseudoscalar is
$x'\bar x$, and the current is  $\bar y\gamma_\mu x$.  The corresponding neutral valence mesons
$x\bar x$, $x'\bar x'$, and $y\bar y$, 
are named $X$, $X'$, and $Y$, respectively.


At the meson level, the vector current
in continuum partially quenched chiral perturbation theory is
\begin{equation}\eqn{Vmu-cont}
V^\mu_{xy} = \frac{if^2}{4}\left[\partial^\mu\Sigma\;\Sigma^\dagger-\Sigma^\dagger\partial^\mu\Sigma
\right]_{xy}\ .
\end{equation}
As always, the left index of $\Sigma$ or $\Sigma^\dagger$ is a quark index, and the right index is an
antiquark index.  For \schpt, we must choose the  taste structure of the current.  Since we want
the current to be diagonal in both taste and flavor when $x=y$, in order that it be related to the 
quark number current and the Ademollo-Gatto theorem apply, we need to choose the singlet-taste current.  
Furthermore, a singlet-taste current will allow the incoming and outgoing mesons both to be pseudoscalar taste, which is by far the easiest choice for simulations.  We thus take
\begin{equation}\eqn{Vmu}
V^\mu_{xy} = \frac{if^2}{4}\;\trt\!\left[\partial^\mu\Sigma\;\Sigma^\dagger-\Sigma^\dagger\partial^\mu\Sigma
\right]_{xy}\ ,
\end{equation}
where $\trt$ is the trace over taste indices only.  

We may check the normalization in \eq{Vmu} by computing the $K$--$\pi$ matrix element at leading order (LO) in \schpt.  In the unitary case ($y=s$, $x=u$, $x'=d$), the relevant two-meson term in the current is
\begin{equation}\eqn{Vmu-LO}
i \sum_\Xi \left(\partial^\mu \pi^{+}_{\Xi}\, K_{\Xi}^0 - \pi^{+}_{\Xi}\,\partial^\mu  K_{\Xi}^0 
\right) \ , 
\end{equation}
with $\Xi$ the taste of the mesons.  The LO matrix element for pseudoscalar-taste mesons is then
\begin{equation}
\langle \pi^-_{5} (p_\pi)\vert\; V_\mu^{us}\;\vert K^0_{5}(p_K)\rangle = p_K^\mu + p_\pi^\mu \ , 
\end{equation}
where the subscript 5 indicates the pseudoscalar taste, that is $\Xi=\xi_5$. 
Thus, $f_+(q^2)=1$ at LO, consistent with the Ademollo-Gatto theorem.

In the following subsections we collect the relevant one-loop S$\chi$PT 
formulae for the extrapolation of the form factor $f_+(q^2)$.  We begin with the simplest contribution,
wave function renormalization.

\subsection{Wave function renormalization}

The one-loop contribution to the wave function renormalization of a pseudoscalar meson $P_{xy}$ (with pseudo scalar taste) can be taken from 
\rcite{SCHPT} or from  Appendix A of \rcite{BA2007}.  We first write it in the compact notation
using \eq{Adisc-value}, as well as \eqs{Avalue}{ell}, for the chiral logarithm function $\ell$:
\begin{eqnarray}\label{eq:ChPTWFR}
  Z_{P_{xy}} & =&\frac{1}{12(4\pi f)^2} \sum_\Xi \Biggl\{
  \frac{1}{4}\sum_{\mathscr{S}}
  \left[
  \ell\left(m^2_{x\mathscr{S},\Xi}\right)
  +\ell\left(m^2_{y\mathscr{S},\Xi}\right)
  \right]
  \nonumber\\&&{}
  +\ell(D_{XX}^\Xi)+\ell(D_{YY}^\Xi) -2c_\Xi\ell(D_{XY}^\Xi)
    \Biggr\}\ , \
\end{eqnarray}
where $\Xi$ runs over the
16 tastes, and  $\mathscr{S}$ runs over the sea quarks ($u,d,s$).  The disconnected propagators 
$D^\Xi$ are given by \eq{DiscXi} (or the simplified form of \eq{DiscI} for the singlet-taste channel).
The rooting procedure, which multiplies the terms involving $\mathscr{S}$ by a factor of $1/4$, and modifies the
sea-meson masses entering into the denominators of the disconnected propagators,
has already been implemented.
The sign factor $c_\Xi$, which arises from commuting the pseudoscalar-taste external fields past the
taste-$\Xi$ internal fields, is defined by
\begin{equation}\eqn{cdef}
c_\Xi = \frac{1}{4}{\rm Tr}\left(\xi_5 \xi_\Xi \xi_5 \xi_\Xi\right)\ .
\end{equation}
Note that the taste of the external meson only enters through these factors; for an external taste other
than pseudoscalar, one merely needs to replace each explicit $\xi_5$ matrix in \eq{cdef} with the 
appropriate  taste matrix.

We may also express the disconnected propagators as sums
 over simple poles times the residue functions $R_j^{[n,k]}$ defined in 
Ref.~\cite{SCHPT}, and write the result in \eq{ChPTWFR} more explicitly in terms of the
chiral logarithm function $\ell(m^2)$, \eq{ell}, only. 
In the  $N_f=1+1+1$ case (\ie no degeneracies in the sea), we have 
\begin{eqnarray}\label{eq:WFR111}
  Z_{P_{xy}} & =&\frac{1}{3(4\pi f)^2}\Biggl\{
  \frac{1}{16}\sum_{\mathscr{S},\Xi}
  \left[
  \ell\left(m^2_{x\mathscr{S},\Xi}\right)
  +\ell\left(m^2_{y\mathscr{S},\Xi}\right)
  \right]
  \nonumber\\&&{}
  +\frac{1}{3}\Biggl[\sum_{j\in\cM^{(3,x)}}
    \frac{\partial}{\partial m_{X,I}^2}
  \left(
  R^{[3,3]}_{j} \left(\cM^{(3,x)}_I ; \mu^{(3)}_I\right)
  \ell(m^2_{j,I})\right)\nonumber \\* &&
  +\sum_{j\in\cM^{(3,y)}}
    \frac{\partial}{\partial m_{Y,I}^2}
    \left(
    R^{[3,3]}_{j}  \left(\cM^{(3,y)}_I ; \mu^{(3)}_I\right)
    \ell(m^2_{j,I})\right)\nonumber\\&&{}
  +2\sum_{j\in\cM^{(4,x,y)}}
  R^{[4,3]}_{j} \left(\cM^{(4,x,y)}_I ; \mu^{(3)}_I\right)
  \ell(m^2_{j,I})\Biggr]
  \nonumber \\* &&{}
+ a^2 \delta_V
  \Biggl[
    \sum_{j\in\cM^{(4,x)}}
    \frac{\partial}{\partial m_{X,V}^2}
  \left(
  R^{[4,3]}_{j}  \left(\cM^{(4,x)}_V ; \mu^{(3)}_V\right)
  \ell(m^2_{j,V})
  \right)\nonumber \\* &&
  +\sum_{j\in\cM^{(4,y)}}
    \frac{\partial}{\partial m_{Y,V}^2}
    \left(
    R^{[4,3]}_{j}\left(\cM^{(4,y)}_V ; \mu^{(3)}_V\right)
    \ell(m^2_{j,V})\right)\nonumber\\&&{}
  -2\sum_{j\in\cM^{(5,x,y)}}
  R^{[5,3]}_{j}\left(\cM^{(5,x,y)}_V ; \mu^{(3)}_V\right)
  \ell(m^2_{j,V})
    \Biggr]
  + \Bigl[ V \to A \Bigr]\Biggr\}\ , \
\end{eqnarray}
Here the derivatives with respect to $m_X^2$ and $m_Y^2$ (of various tastes) arise from the
double poles in the disconnected propagators $D_{XX}^\Xi$ and $D_{YY}^\Xi$. 
The arguments $\cM$ and $\mu$ of the residue functions $R_j^{[n,k]}$ 
are various sets of meson masses:
\begin{eqnarray}\label{eq:denom_mass_sets}
        \{\cM^{(3,z)}_\Xi\}& \equiv & \{m_{\pi^0,\Xi},\,m_{\eta,\Xi},\,
                m_{Z,\Xi}   \}\ , \nonumber \\*
        \{\cM^{(4,z,z')}_\Xi\}& \equiv & \{m_{\pi^0,\Xi},\,m_{\eta,\Xi},\,
        m_{Z,\Xi},\,m_{Z',\Xi}\}\ , \nonumber \\*
        \{\cM^{(4,z)}_\Xi\}& \equiv & \{m_{\pi^0,\Xi},\,m_{\eta,\Xi},\,m_{\eta',\Xi},\,
                m_{Z,\Xi}   \}\ , \nonumber \\*
        \{\cM^{(5,z,z')}_\Xi\}& \equiv & \{m_{\pi^0,\Xi},\,m_{\eta,\Xi},\,m_{\eta',\Xi},\,
        m_{Z,\Xi},\,m_{Z',\Xi}\}\ , \\*
        \{\mu^{(3)}_\Xi\}& \equiv & \{m_{U,\Xi},\,m_{D,\Xi},\,m_{S,\Xi}
                \}\ , \nonumber
\end{eqnarray}
where $z$ and $z'$ can be any valence quark ($x$, $x'$ or $y$), and $Z$ and $Z'$ are the corresponding
$z\bar z$ or $z' \bar z'$ mesons ($X$, $X'$, or $Y$).  
For 
our partially quenched version of $K\to\pi l \nu$, we need $(Z_{P_{xx'}}+Z_{P_{yx'}})/2$.

In the 2+1 case, the $\pi$ is degenerate with the diagonal $U$ and $D$ states, so $\pi$
should be eliminated from the denominator sets $\cM$, $D$ (say) should be
eliminated from the numerator set $\mu$, and each index of each residue function
should be reduced by $1$:
$$R_j^{[n,k]} \to R_j^{[n-1,k-1]}\ . $$

\subsection{Strong vertex diagram}

Aside from wave function renormalization, there are two chiral diagrams 
that contribute to the form factors $f_+(q^2)$ and $f_-(q^2)$, which are shown in
\figref{meson-diagrams}.  Diagram \figref{meson-diagrams}(i) contains 
an $\order(p^0)$ $\Delta S=1$ current vertex involving
two fields and a strong $\order(p^2)$ vertex involving four fields. We call it 
the ``strong vertex'' diagram, to distinguish it from \figref{meson-diagrams}(ii), 
which contains only a current vertex.
A significant simplification to the evaluation of the strong vertex diagram comes from 
the fact that we calculate only the form factor $f_+$.  The derivative
$\partial^\mu$ in the current vertex introduces  a factor of $k^\mu$ or $(k-q)^\mu$, where $k$ is the loop 
momentum.  Since the loop integrand depends only on $k$ and $q$, in most cases the integration over
$k$ will give a result proportional to $q^\mu$, so the diagram becomes a contribution to $f_-$, not $f_+$.
For example, this will occur when the strong vertex in the graph comes from the mass term in the 
Lagrangian, Eq.~(\ref{LSChPT}). The only exception occurs when the vertex comes from the kinetic energy
term, with one of its derivatives acting on an internal line and the other acting on an external line. This
can introduce a factor of $k^\nu p^\nu$ where $p$ is either $p_K$ or $p_\pi$.  The integration over
$k$ will then produce a term proportional to $\delta^{\mu\nu}$ (see \eq{C22-def}). There is thus a
contribution that goes like $p^\mu$ and hence can contribute to $f_+$ (as well as to $f_-$).

\begin{figure}[tb]

\vspace*{-2.cm}

\begin{minipage}[c]{0.49\textwidth}

\hspace*{-1.cm}\includegraphics[width=1.2\textwidth]{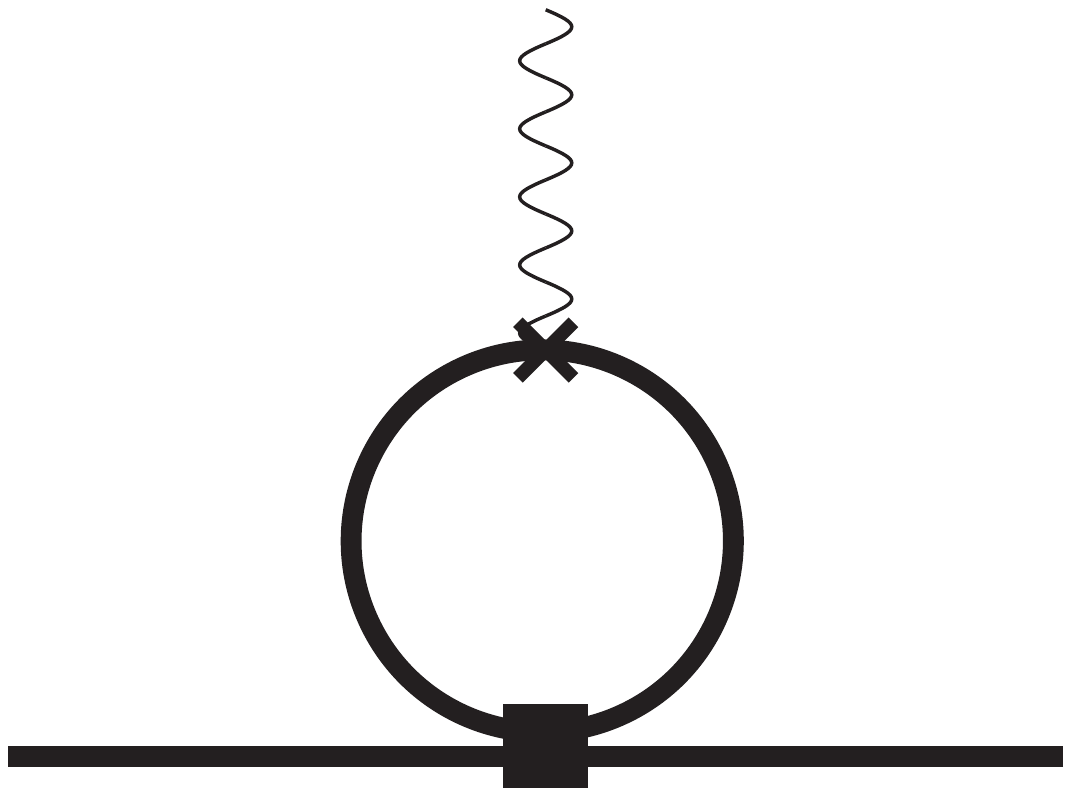}

\vspace*{-8.5cm}
\begin{center}{\bf (i)}\end{center}
\end{minipage}
\begin{minipage}[c]{0.49\textwidth}

\hspace*{-1.cm}\includegraphics[width=1.2\textwidth]{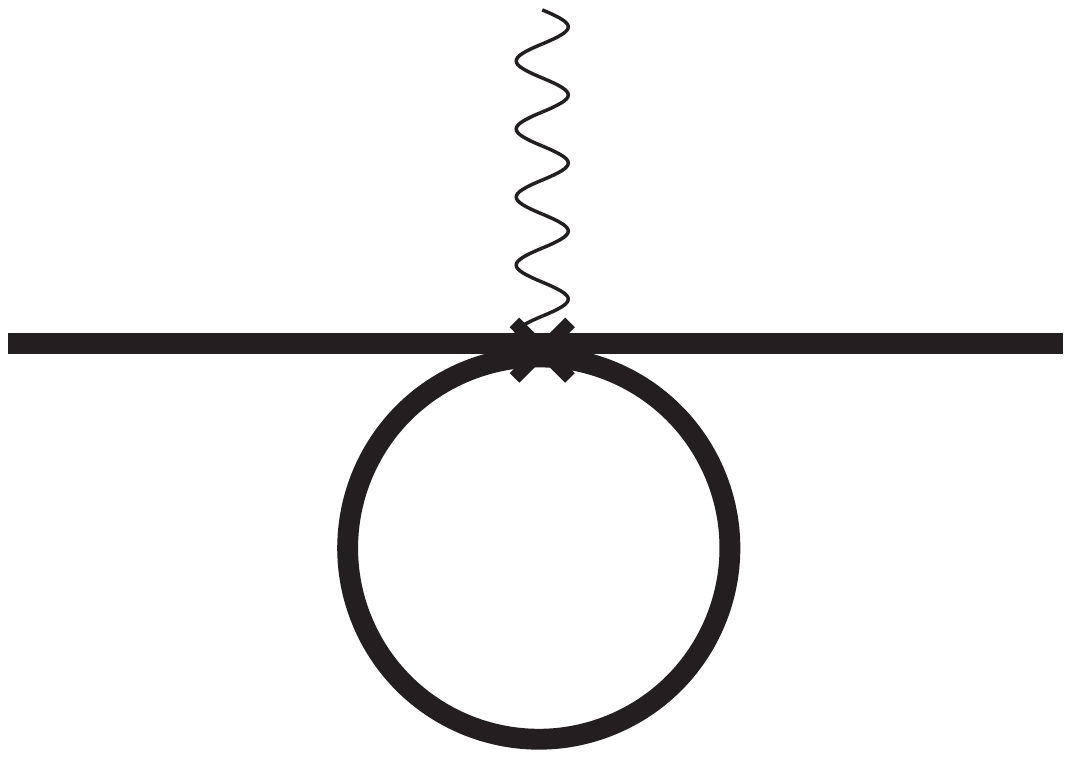}

\vspace*{-8.5cm}
\begin{center}{\bf (ii)}\end{center}
\end{minipage}
\caption{\label{fig:meson-diagrams}Meson diagrams contributing to $f_+(q^2)$. The bold 
${\bm \times}$ with attached wavy line represents the weak vector current 
and the filled squared represents a strong vertex. 
Diagram (i) consists of the two-meson
contribution to the current, together with a strong 4-meson vertex from the Lagrangian.  
Diagram (ii) has a four-meson contribution to the current.}
\end{figure}

There are seven different topologies of this type. One, shown in \figref{ChPT0}, has  
all connected internal meson propagators, while six, shown in Fig.~\ref{fig:ChPT1}, 
have a disconnected meson propagator.

\begin{figure}[tb]

\vspace*{-1.cm}

\begin{minipage}[c]{0.49\textwidth}
\includegraphics[width=0.95\textwidth]{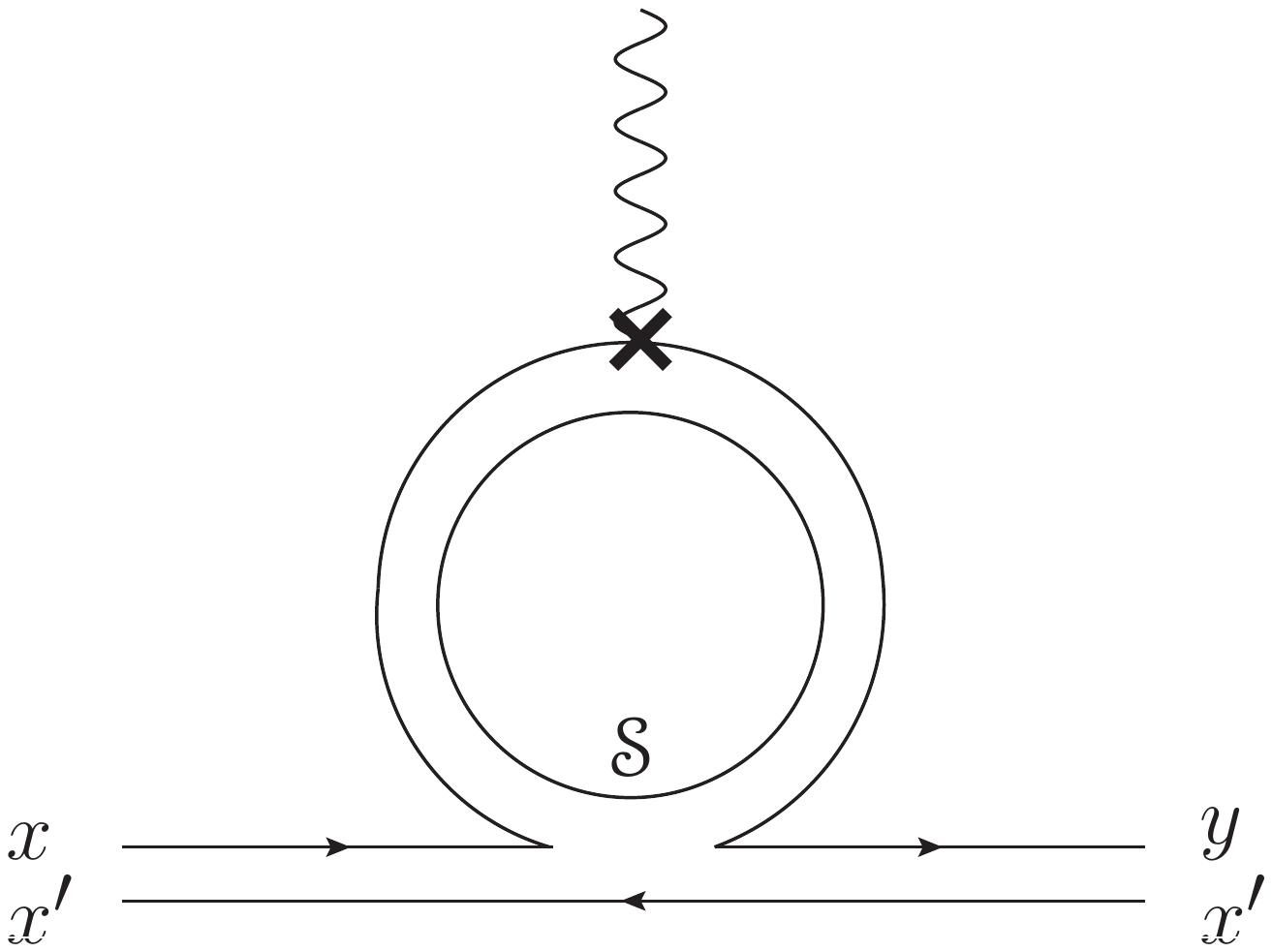}

\vspace*{-6.5cm}
\begin{center}{\bf (a)}\end{center}
\end{minipage}
\caption{\label{fig:ChPT0}Quark flow for the strong vertex diagram with connected 
internal meson propagators. }
\end{figure}

\begin{figure}[tb]

\vspace*{-2.cm}

\begin{minipage}[c]{0.49\textwidth}

\includegraphics[width=0.95\textwidth]{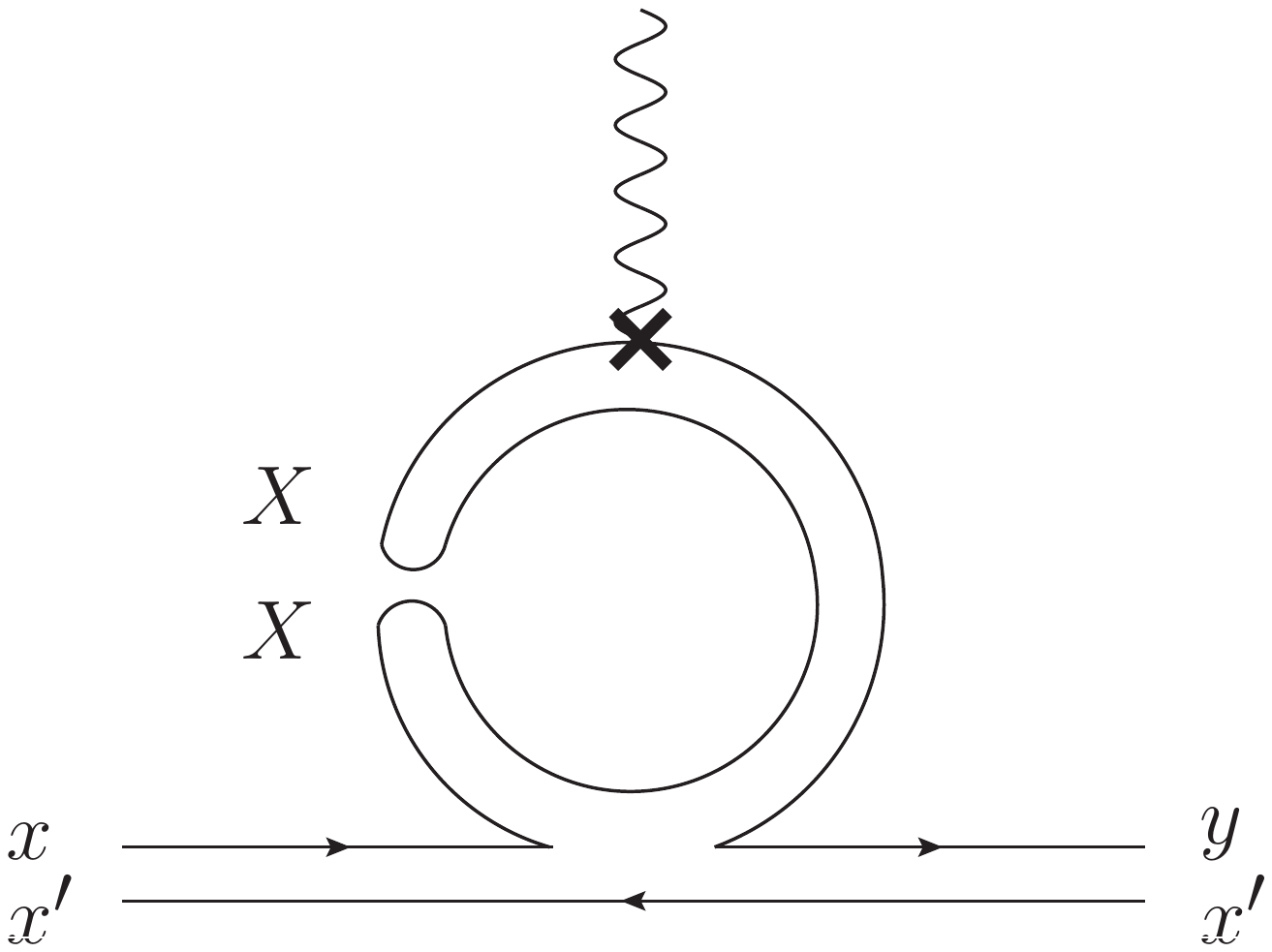}

\vspace*{-6.5cm}
\begin{center}{\bf (b)}\end{center}
\end{minipage}
\begin{minipage}[c]{0.49\textwidth}
\includegraphics[width=0.95\textwidth]{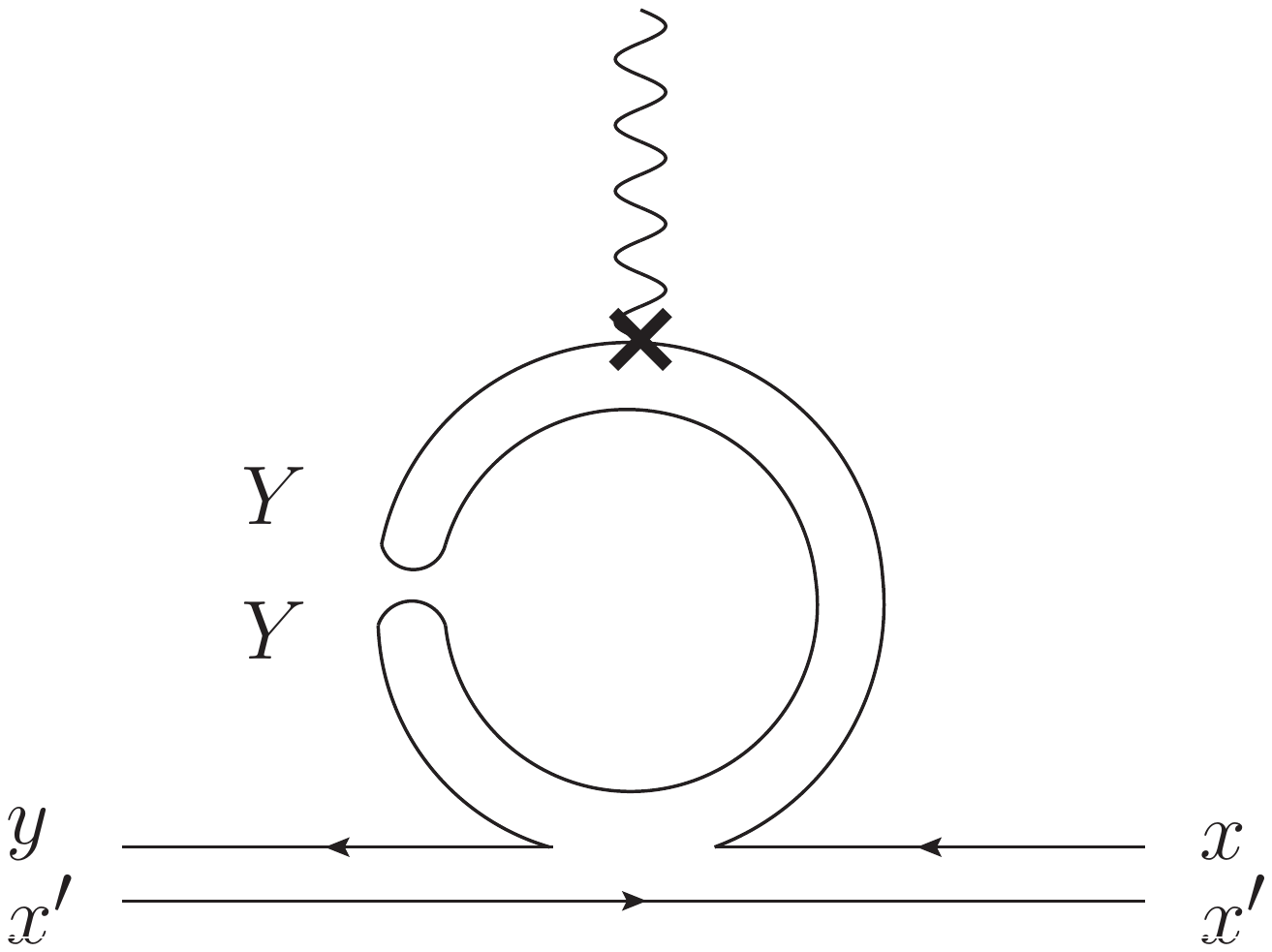}

\vspace*{-6.5cm}
\begin{center}{\bf (c)}\end{center}
\end{minipage}
\begin{minipage}[c]{0.49\textwidth}

\vspace*{-0.2cm}
\includegraphics[width=0.95\textwidth]{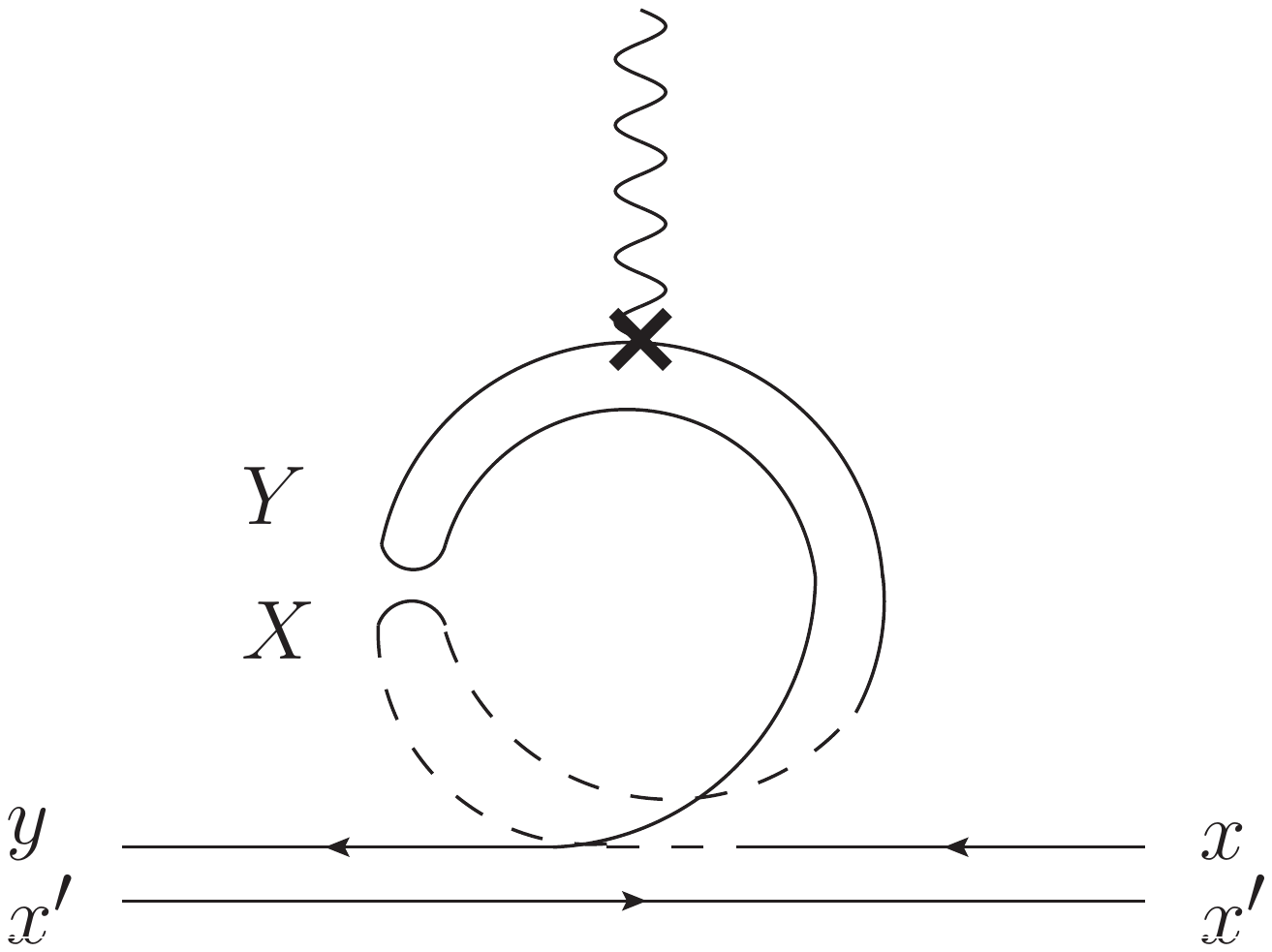}

\vspace*{-6.5cm}
\begin{center}{\bf (d)}\end{center}
\end{minipage}
\begin{minipage}[c]{0.49\textwidth}

\vspace*{-0.2cm}
\includegraphics[width=0.95\textwidth]{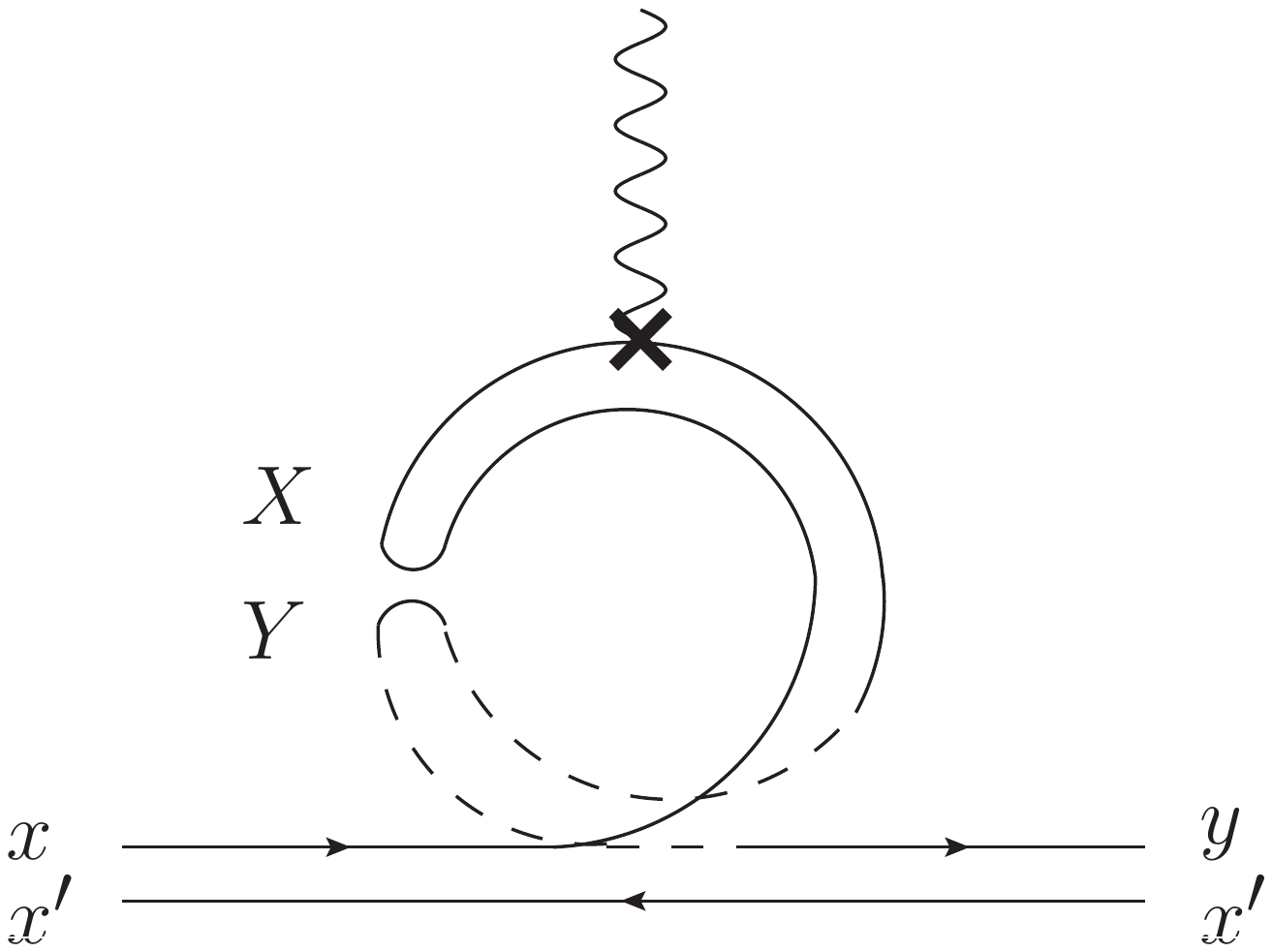}

\vspace*{-6.5cm}
\begin{center}{\bf (e)}\end{center}
\end{minipage}
\begin{minipage}[c]{0.49\textwidth}

\vspace*{-0.2cm}
\includegraphics[width=0.95\textwidth]{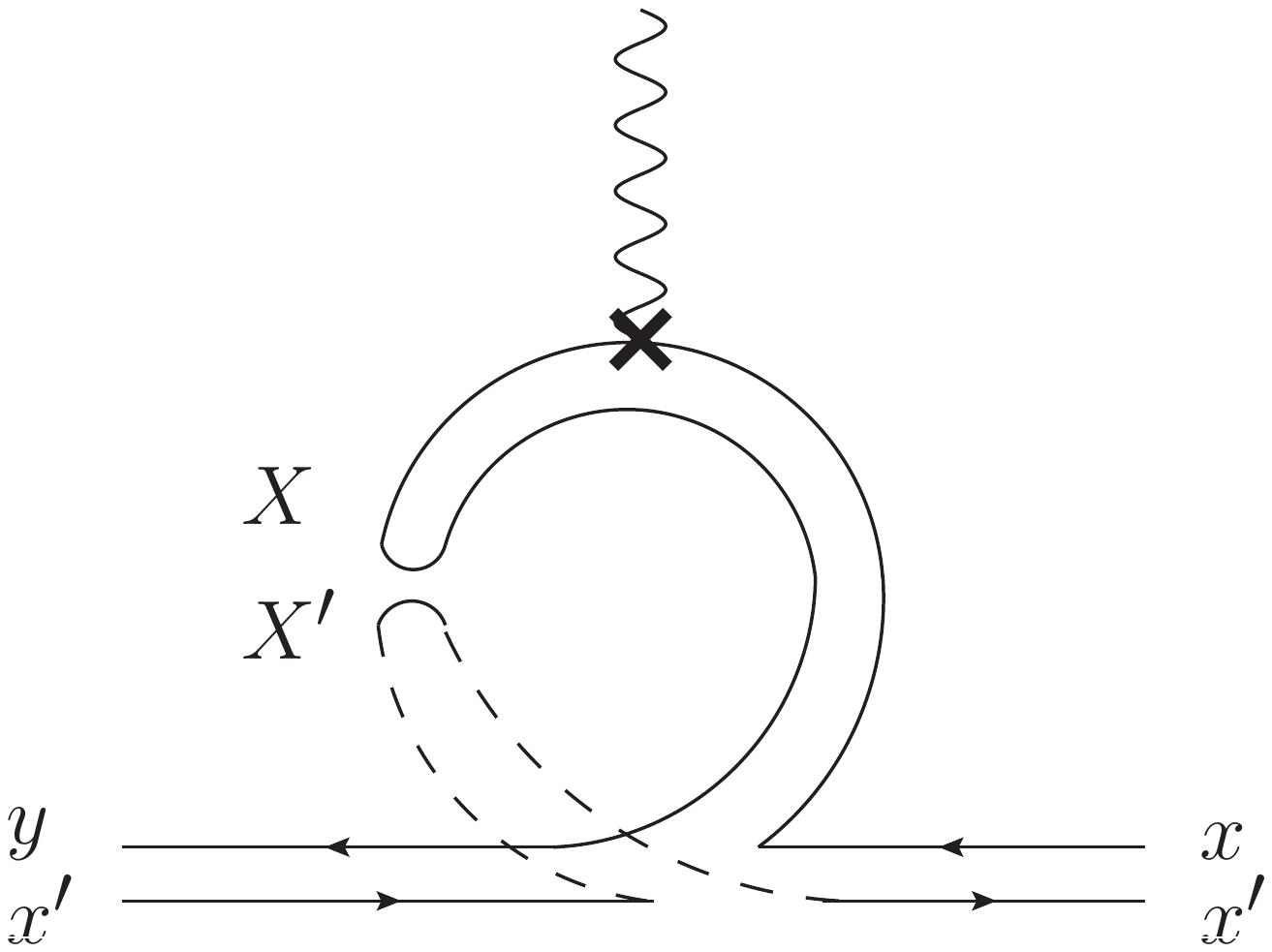}

\vspace*{-6.5cm}
\begin{center}{\bf (f)}\end{center}
\end{minipage}
\begin{minipage}[c]{0.49\textwidth}

\vspace*{-0.2cm}
\includegraphics[width=0.95\textwidth]{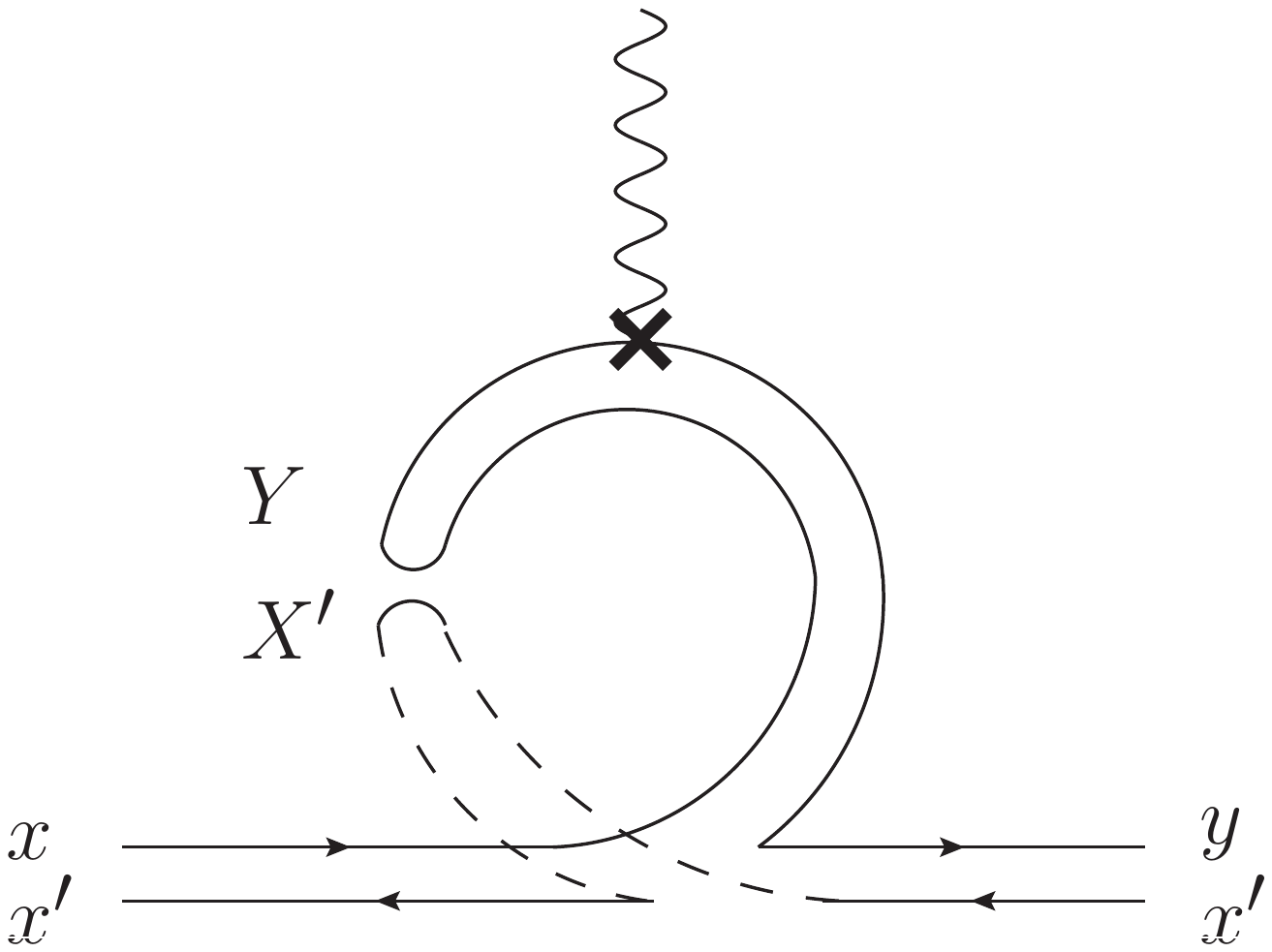}

\vspace*{-6.5cm}
\begin{center}{\bf (g)}\end{center}
\end{minipage}
\caption{\label{fig:ChPT1} Quark flow for the strong vertex diagrams.}
\end{figure}

In the compact notation of  \eq{Cdisc-value}, as well as \eqs{Cvalue}{B22-value}, 
the strong vertex diagrams give the following contribution to $f_+(q^2)$:  
\vspace*{-0.2cm}
\begin{eqnarray}
\frac{1}{2(4\pi f)^2} \sum_\Xi \Biggl\{  -\frac{1}{4}\sum_{\mathscr{S}} &
\hspace{-2mm}\tilde B_{22}(m^2_{x\mathscr{S},\Xi},m^2_{y\mathscr{S},\Xi},q^2)
&\quad{\bf (a)}\nonumber\\ 
& \hspace{-8mm}- \tilde B_{22}(m_{xy,\Xi}^2,D^\Xi_{XX},q^2) 
 &\quad {\bf (b)} \nonumber\\ 
&\hspace{-8mm} - \tilde B_{22}(m_{xy,\Xi}^2,D^\Xi_{YY},q^2) 
 & \quad {\bf (c)} \nonumber\\ 
&\hspace{-8mm} + 2 \tilde B_{22}(m_{xy,\Xi}^2,D^\Xi_{XY},q^2) 
& \quad {\bf (d)+(e)}   \Biggl\} \, ,\label{eq:b22-compact}
\end{eqnarray}
where the bold letter(s) after each term indicate(s) the diagram(s) it comes from. 
Note that $\tilde B_{22}$  
is a factor of $(4\pi)^2$ larger than the corresponding function  $\bar B_{22}$ 
defined in \rcite{Bijnens:2002hp}.  Diagrams {\bf f} and {\bf g} do not contribute to $f_+(q^2)$ because the strong
vertex gives no terms with  the needed single factor of the loop momentum $k$; those diagrams 
do however contribute to $f_-(q^2)$.  Since the only strong vertex diagrams where the spectator quark
plays a role are {\bf f} and {\bf g}, \eq{b22-compact} is independent of the spectator 
quark $x'$.

Explicitly, in the $N_f=1+1+1$ case, \eq{b22-compact} becomes
\vspace*{-0.2cm}
\begin{eqnarray}
\frac{2}{(4\pi f)^2} \Biggl\{ && -\frac{1}{16}\sum_{\mathscr{S},\Xi} \, 
\tilde B_{22}(m^2_{x\mathscr{S},\Xi},m^2_{y\mathscr{S},\Xi},q^2)\quad 
{\bf (a)}\nonumber\\ 
&& + \frac{1}{3}\left\lbrack - \frac{\partial}{\partial m_{X,I}^2} \left\lbrace 
\sum_j \tilde B_{22}(m_{xy,I}^2,m_{j,I}^2,q^2) 
R^{[3,3]}_{j} 
\left(\cMI^{(3,x)} ; \mu^{(3)}_I\right)\right\rbrace\
\right\rbrack \quad {\bf (b)} \nonumber\\ 
&& + \frac{1}{3}\left\lbrack - \frac{\partial}{\partial m_{Y,I}^2} \left\lbrace 
\sum_j \tilde B_{22}(m_{xy,I}^2,m_{j,I}^2,q^2) 
R^{[3,3]}_{j} 
\left(\cMI^{(3,y)} ; \mu^{(3)}_I\right)\right\rbrace\
\right\rbrack \quad {\bf (c)} \nonumber\\ 
&&-\frac{2}{3} \left \lbrack \sum_{j}\, 
\tilde B_{22}(m_{xy,I}^2,m_{j,I}^2,q^2) R^{[4,3]}_j \left(\cMI^{(4,x,y)} ; 
\mu^{(3)}_I\right)\right\rbrack \quad {\bf (d)+(e)}\nonumber\\
&& \hspace*{-1.5cm}+ a^2\delta_V \Biggl[ 
\left( - \frac{\partial}{\partial m_{X,V}^2}
\left\lbrace 
\sum_j \tilde B_{22}(m_{xy,V}^2,m_{j,V}^2,q^2) 
R^{[4,3]}_{j} \left(\cMV^{(4,x)} 
; \mu^{(3)}_V\right) \right\rbrace
\right) \quad {\bf (b)} \nonumber\\ 
&& \left( - \frac{\partial}{\partial m_{Y,V}^2}
\left\lbrace 
\sum_j \tilde B_{22}(m_{xy,V}^2,m_{j,V}^2,q^2) 
R^{[4,3]}_{j} \left(\cMV^{(4,y)} 
; \mu^{(3)}_V\right) \right\rbrace
\right) \quad {\bf (c)} \nonumber\\ 
&&-2 \left( \sum_{j}\, 
\tilde B_{22}(m_{xy,V}^2,m_{j,V}^2,q^2) R^{[5,3]}_j \left(\cMV^{(5,x,y)} ; 
\mu^{(3)}_V\right) \right)\quad {\bf (d)+(e)}\,\Biggl\rbrack\nonumber\\
&& +  \Bigl[ V \to A \Bigr]  \Biggl\} \, .\label{eq:b22-terms}
\end{eqnarray}
where again the bold letter(s) after each term indicate(s) the diagram(s) it comes from. 
In Eq.~(\ref{eq:b22-terms}) and below, the sums over $j$ are always over all masses
in the denominator mass sets $\cM$ given in \eq{denom_mass_sets}. 

\subsection{Current vertex diagram}

In this type of diagram there is a $\order(p^2)$ $\Delta S=1$ vertex involving four 
fields. There are three topologies with connected internal mesons, shown in 
Fig.~\ref{fig:ChPT2}, and six with disconnected internal meson
propagators, shown in Fig.~\ref{fig:ChPT3}.

\begin{figure}[tb]

\vspace*{-2.cm}

\begin{minipage}[c]{0.49\textwidth}

\includegraphics[width=0.95\textwidth]{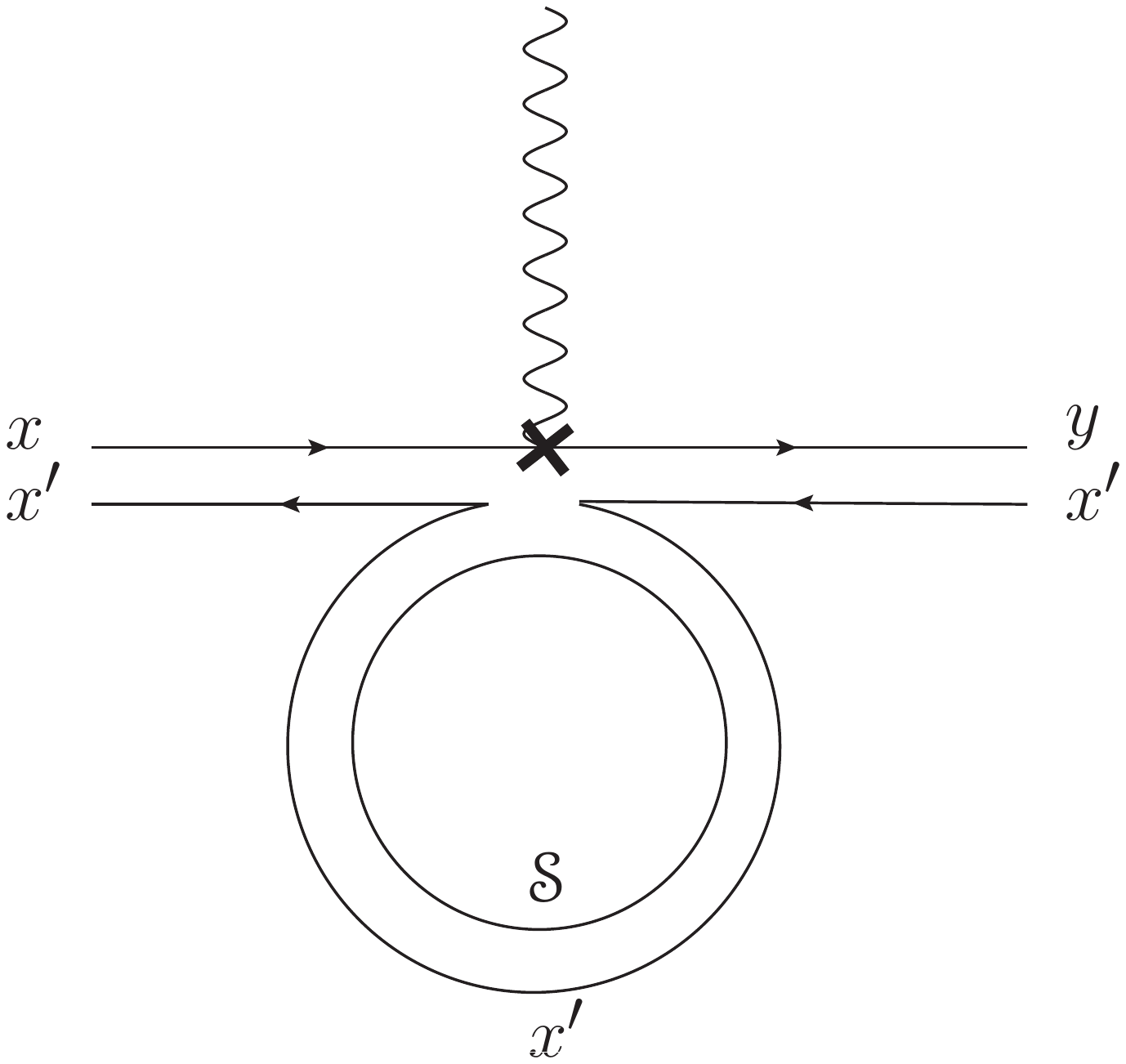}

\vspace*{-5.3cm}
\begin{center}{\bf(h)}\end{center}
\end{minipage}
\begin{minipage}[c]{0.49\textwidth}
\includegraphics[width=0.95\textwidth]{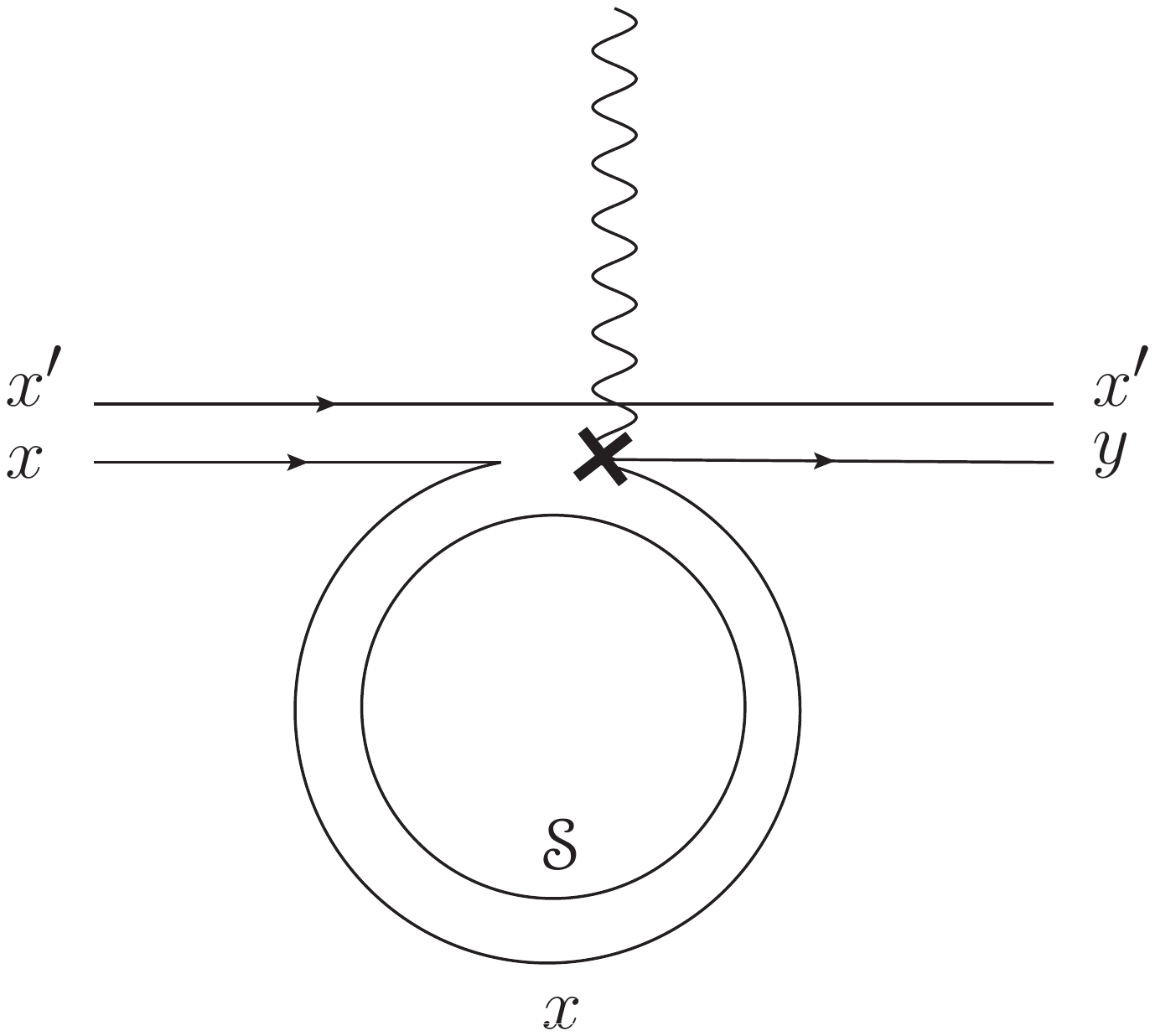}

\vspace*{-5.3cm}
\begin{center}{\bf (\bhp)}\end{center}
\end{minipage}
\begin{minipage}[c]{0.49\textwidth}

\vspace*{-0.5cm}
\includegraphics[width=0.95\textwidth]{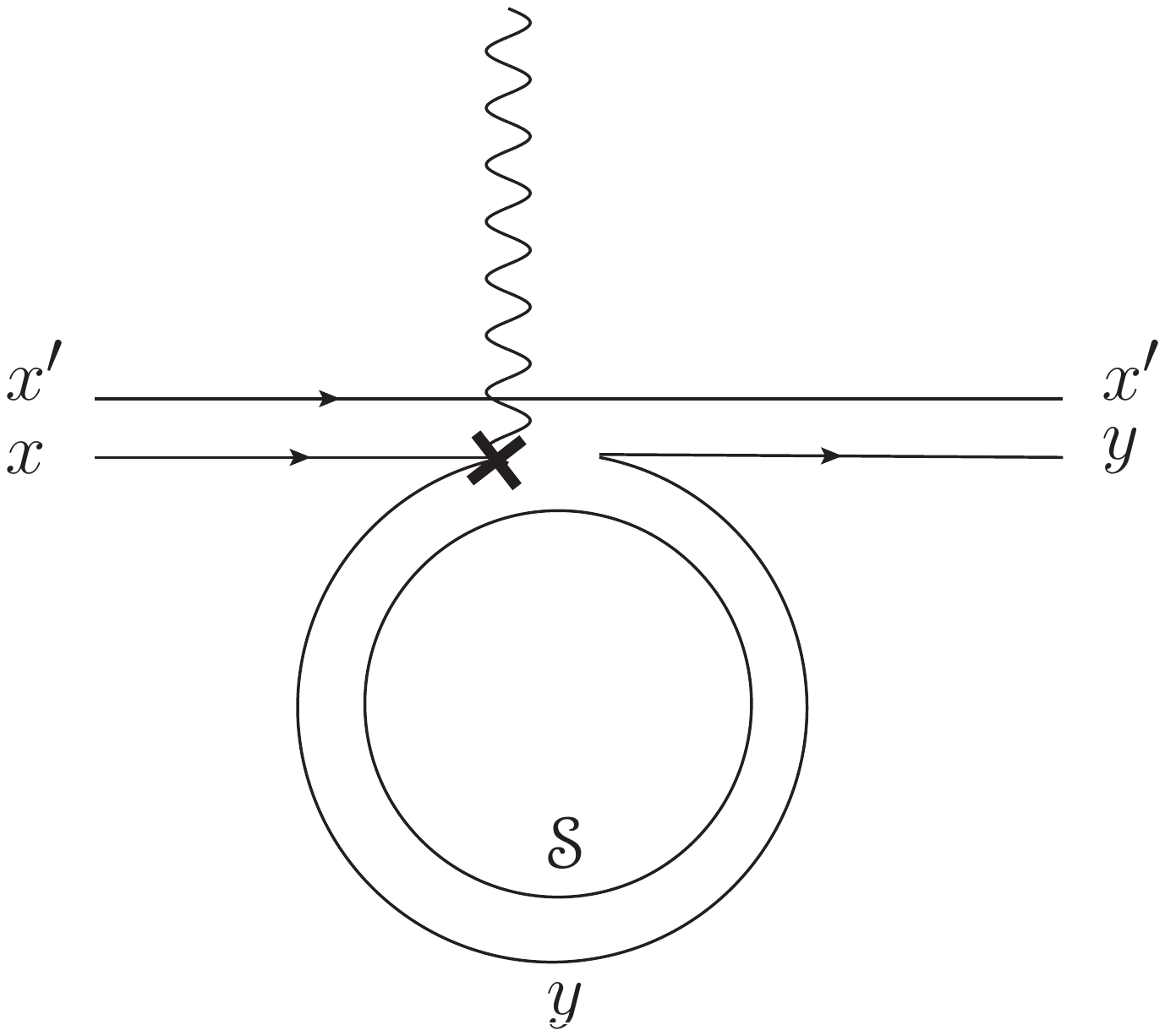}

\vspace*{-5.5cm}
\begin{center}{\bf (\bhpp)}\end{center}
\end{minipage}
\caption{\label{fig:ChPT2}Quark flow for current vertex diagrams that have connected 
internal meson propagators. }
\end{figure}

\begin{figure}[tb]

\vspace*{-2.cm}

\begin{minipage}[c]{0.49\textwidth}

\includegraphics[width=0.95\textwidth]{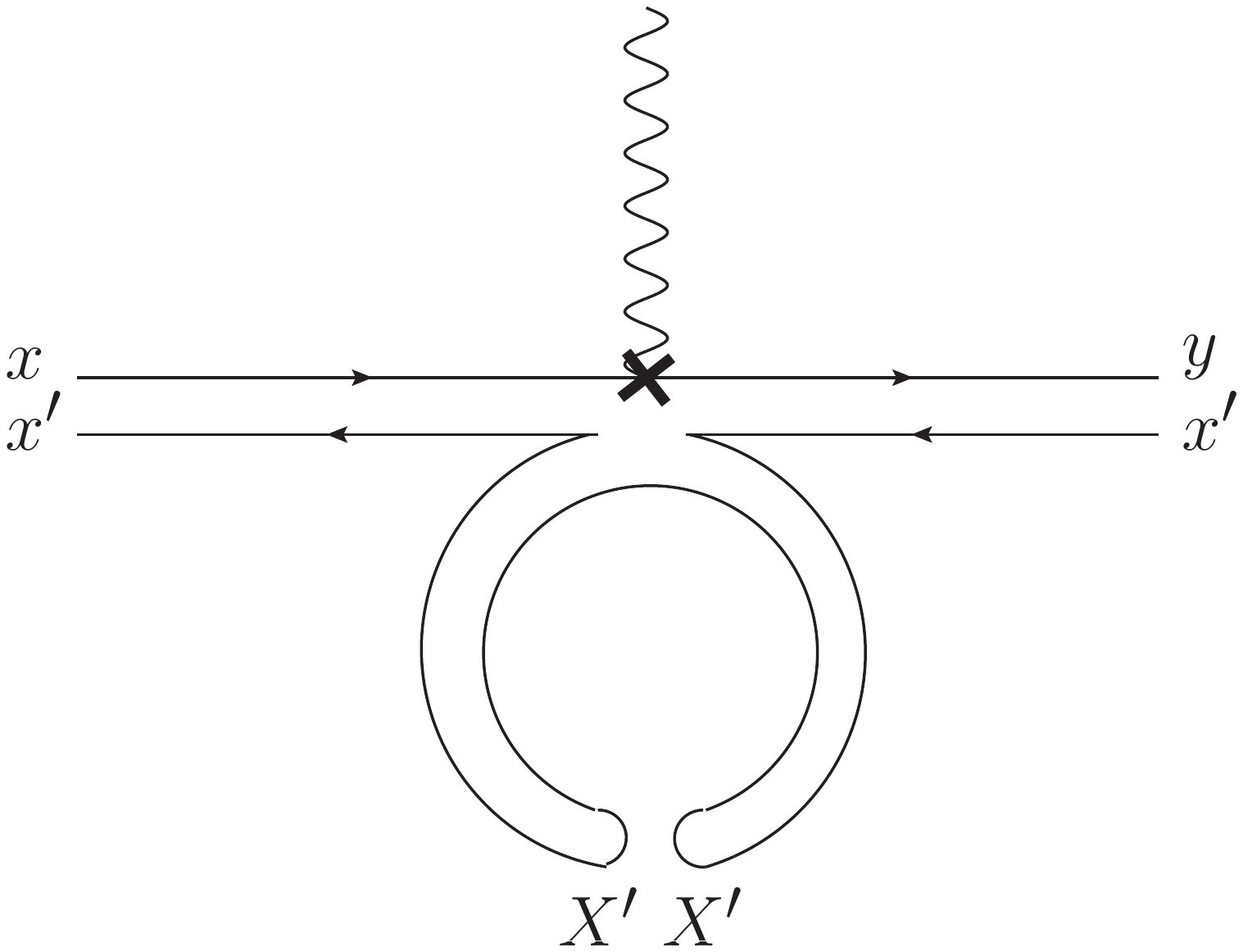}

\vspace*{-5.9cm}
\begin{center}{\bf (j)}\end{center}
\end{minipage}
\begin{minipage}[c]{0.49\textwidth}
\includegraphics[width=0.95\textwidth]{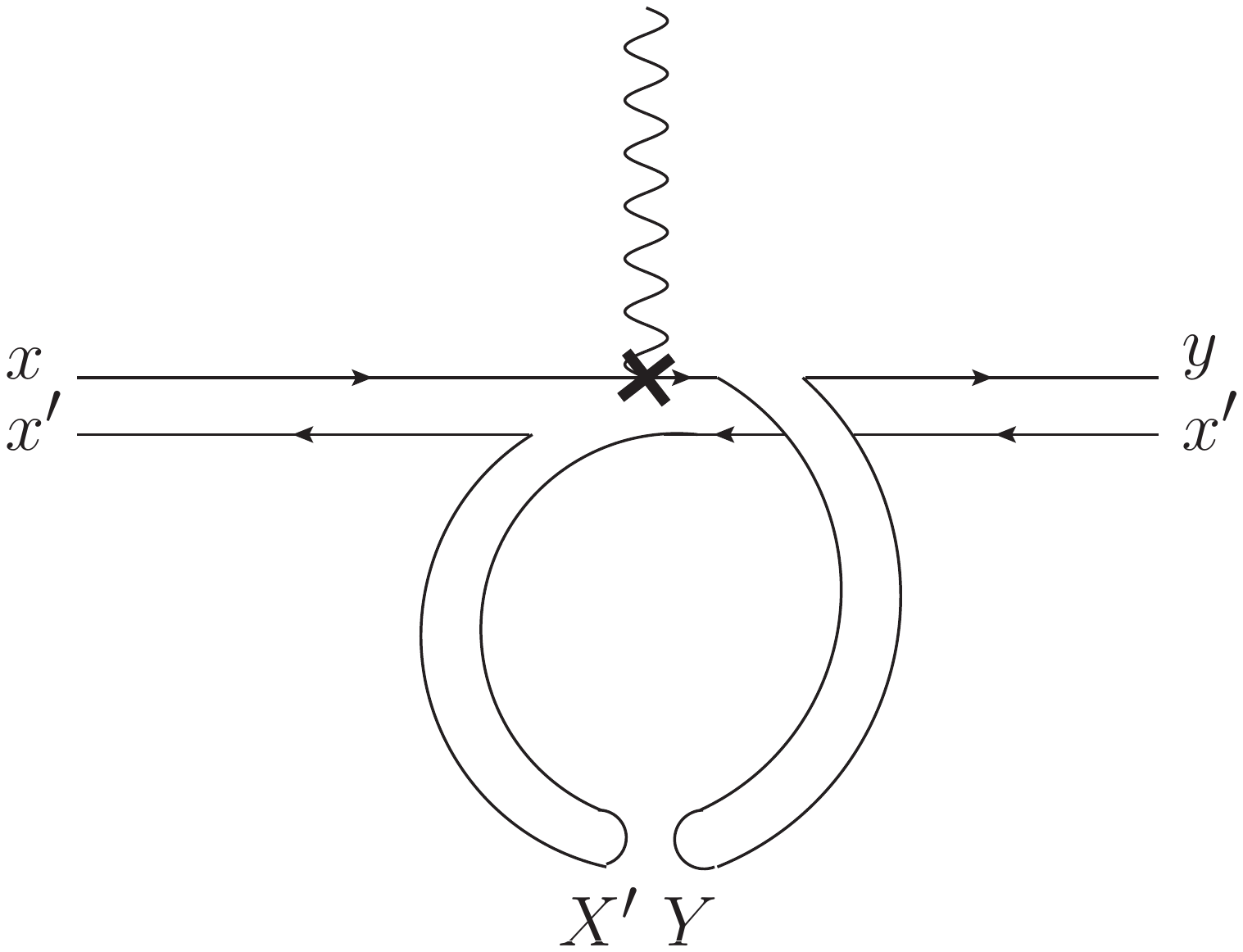}

\vspace*{-5.9cm}
\begin{center}{\bf (k)}\end{center}
\end{minipage}
\begin{minipage}[c]{0.49\textwidth}

\vspace*{-0.5cm}
\includegraphics[width=0.95\textwidth]{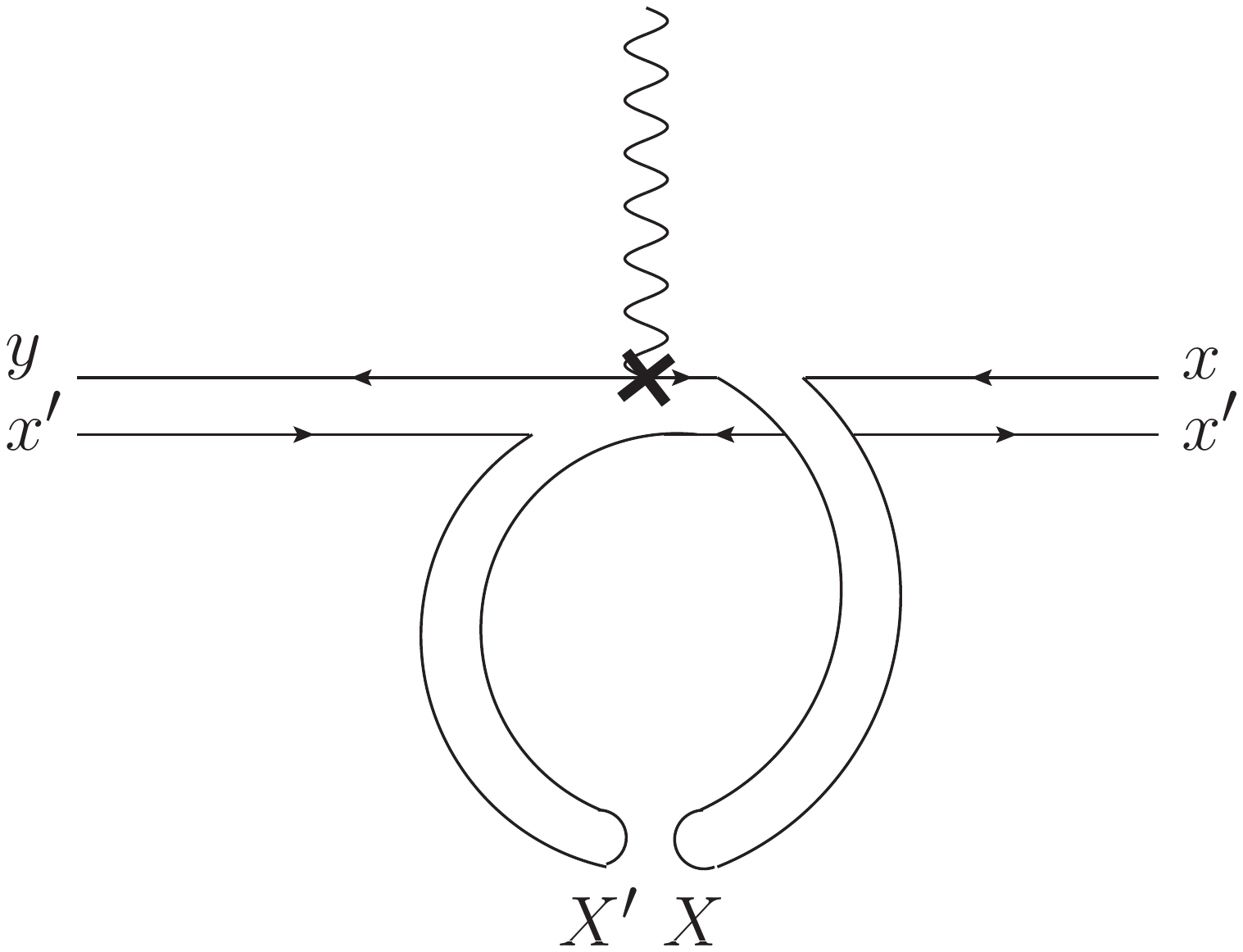}

\vspace*{-5.9cm}
\begin{center}{\bf (l)}\end{center}
\end{minipage}
\begin{minipage}[c]{0.49\textwidth}

\vspace*{-0.5cm}
\includegraphics[width=0.95\textwidth]{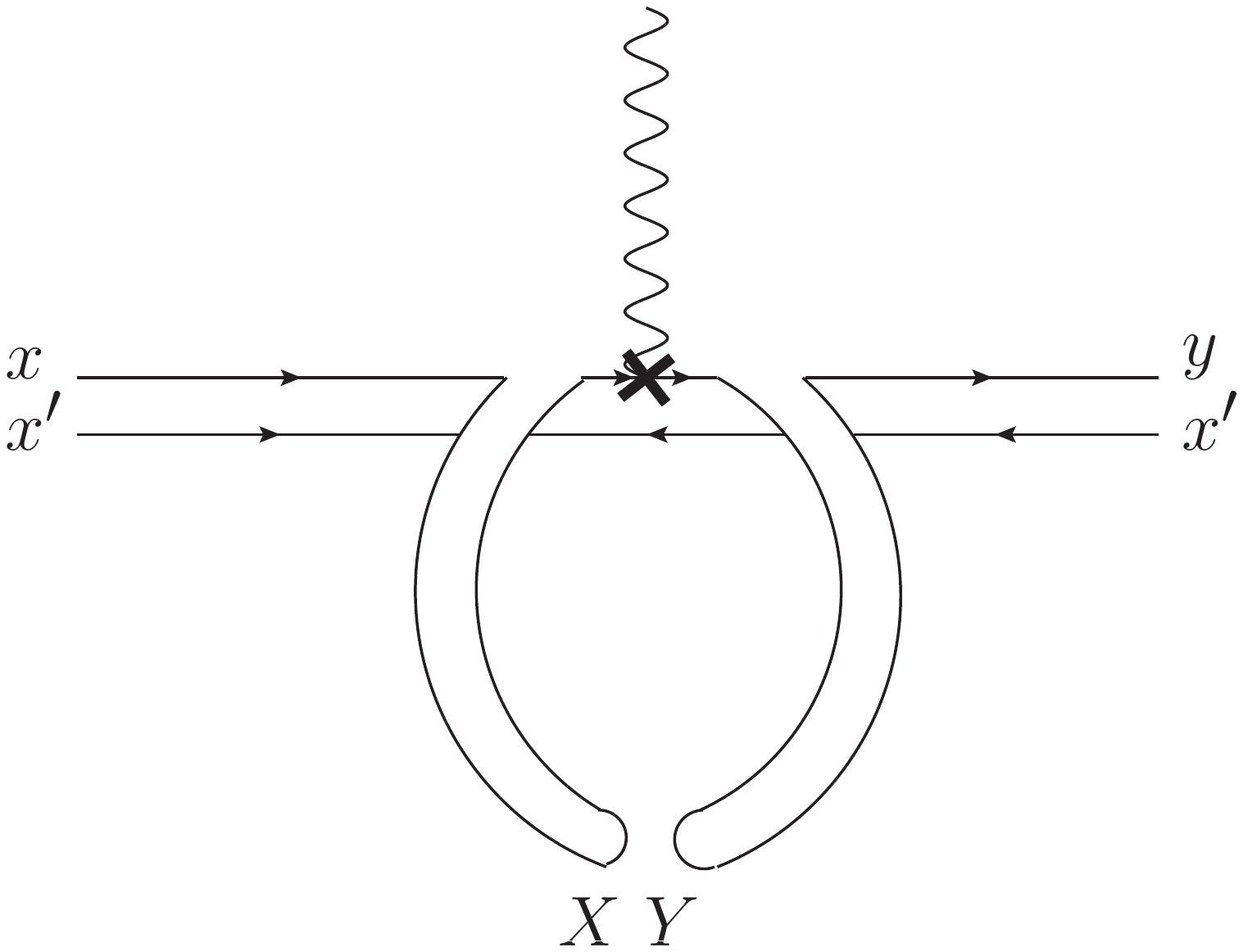}

\vspace*{-5.9cm}
\begin{center}{\bf(m)}\end{center}
\end{minipage}
\begin{minipage}[c]{0.49\textwidth}

\vspace*{-0.5cm}
\includegraphics[width=0.95\textwidth]{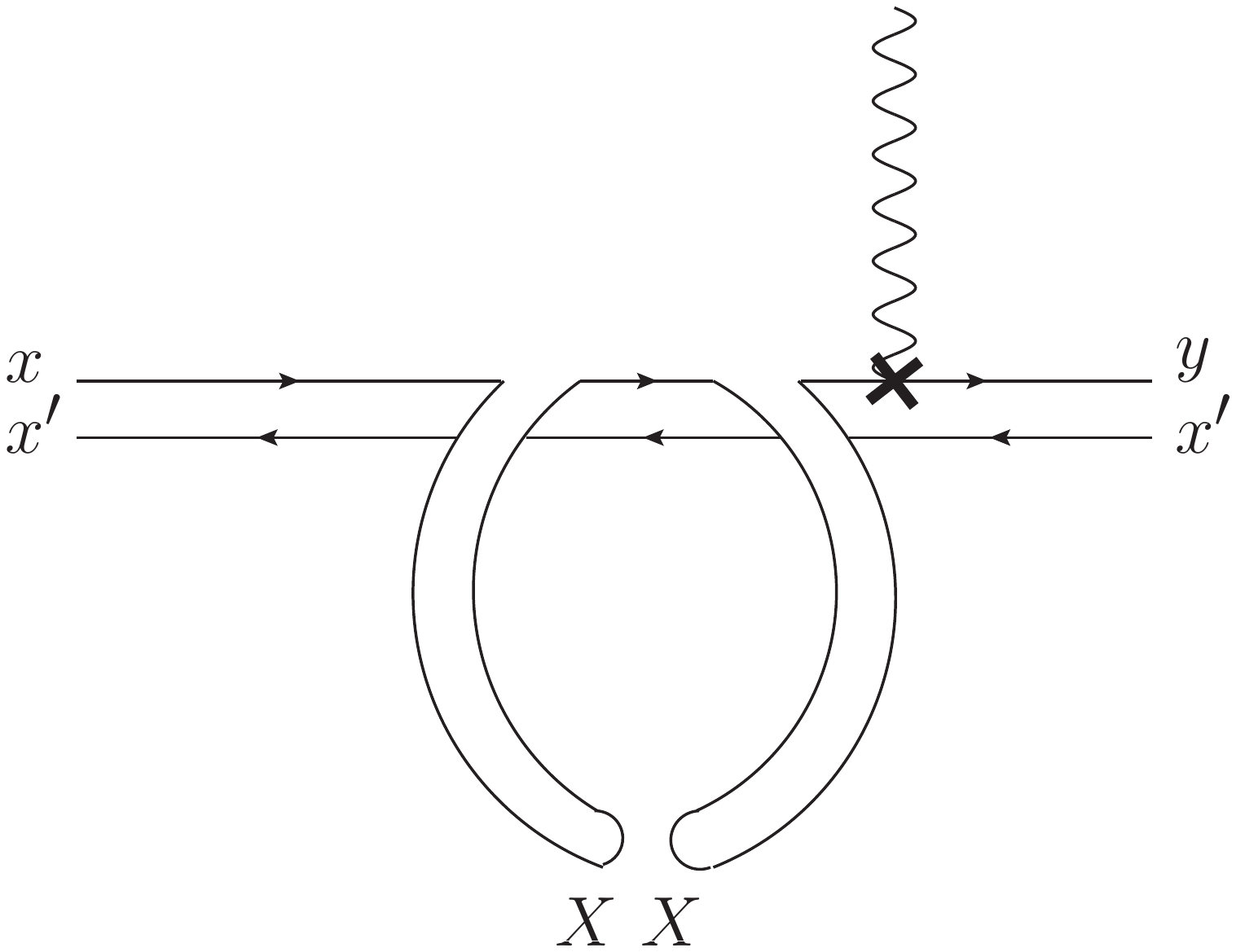}

\vspace*{-5.9cm}
\begin{center}{\bf (n)}\end{center}
\end{minipage}
\begin{minipage}[c]{0.49\textwidth}

\vspace*{-0.5cm}
\includegraphics[width=0.95\textwidth]{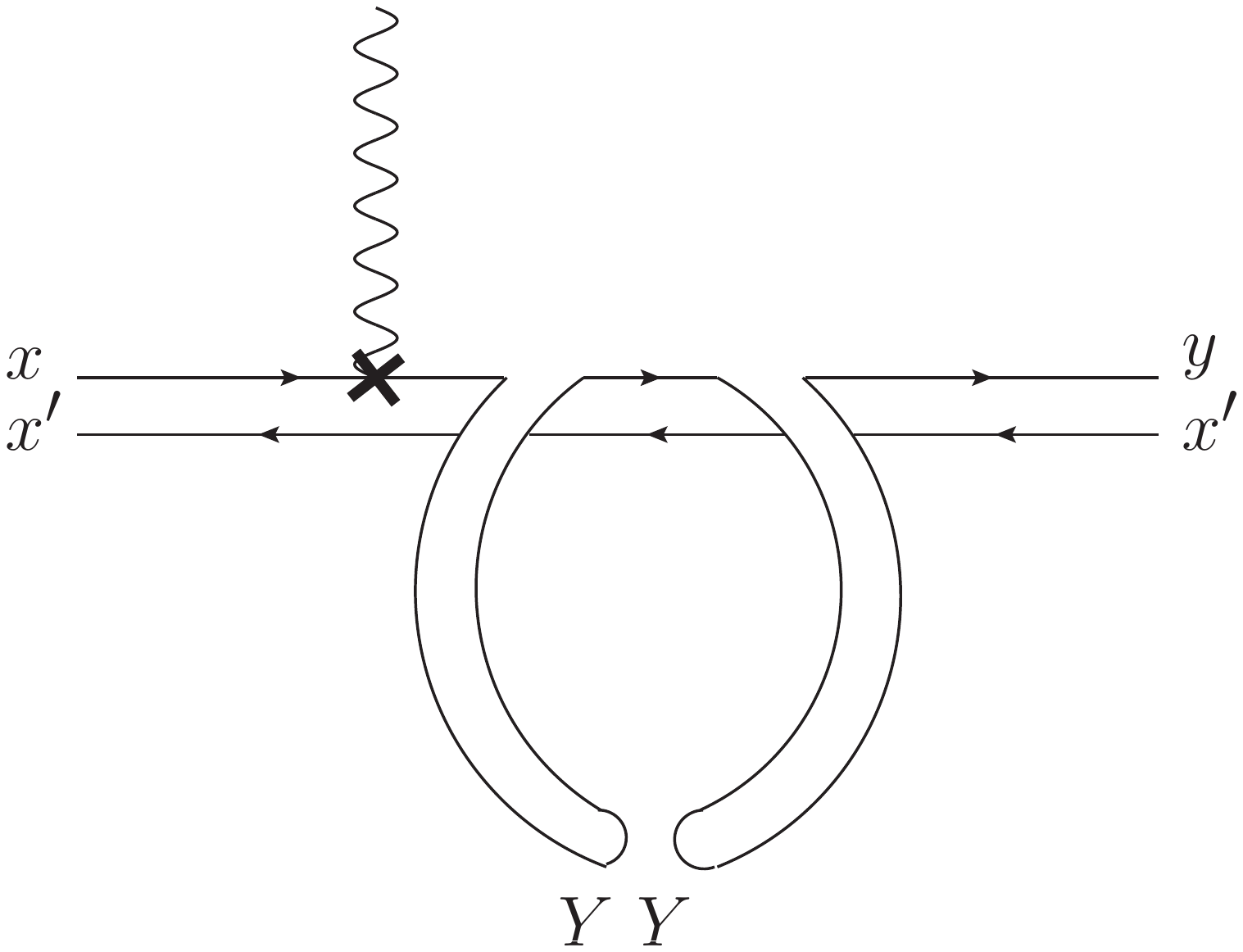}

\vspace*{-5.9cm}
\begin{center}{\bf (o)}\end{center}
\end{minipage}

\caption{\label{fig:ChPT3}Quark flow for current vertex diagrams that have 
disconnected internal meson propagators.} 
\end{figure}

In compact notation, the sum of the contribution from these diagrams to $f_+(q^2)$ is
\begin{eqnarray}
\frac{1}{12(4\pi f)^2} \sum_\Xi \Biggl\{ \frac{1}{4}\sum_{\mathscr{S}}
\Big[-\ell\left(m^2_{x'\mathscr{S},\Xi}\right) & -2\ell\left(m^2_{x\mathscr{S},\Xi}\right)  
-2\ell\left(m^2_{y\mathscr{S},\Xi}\right)\Big] & {\bf (h)+(h')+(h'')} \nonumber \\
 - \ell(D^\Xi_{X'X'}) \qquad{\bf (j)}  &
  \hspace{12mm}- 2 \ell(D^\Xi_{XX})  & {\bf (n)} \nonumber \\
 - 2  \ell(D^\Xi_{YY})  \qquad {\bf (o)} &  
  \hspace{12mm}+ c_\Xi\;\ell(D^\Xi_{XX'})  & {\bf (l)} \nonumber \\
  + c_\Xi\; \ell(D^\Xi_{X'Y})  \qquad {\bf (k)} &  
   \hspace{12mm}+ 3 \ell(D^\Xi_{XY})  & {\bf (m)} \eqn{current-vertex-compact}
\Biggr\}\ .
\end{eqnarray}

In the  $N_f=1+1+1$ case, \eq{current-vertex-compact} becomes, explicitly,
\begin{eqnarray}
\frac{1}{3(4\pi f)^2} \,\Biggl\{ &&\frac{1}{16}\sum_{\mathscr{S},\Xi}
\left[-\ell\left(m^2_{x'\mathscr{S},\Xi}\right) -2\ell\left(m^2_{x\mathscr{S},\Xi}\right)  
-2\ell\left(m^2_{y\mathscr{S},\Xi}\right)\right] \quad {\bf (h)+(h')+(h'')} \nonumber \\
 + \frac{1}{3}\Biggl[ && - \sum_{j\in\cM^{(3,x')}}
    \frac{\partial}{\partial m_{X',I}^2}
  \left(
  R^{[3,3]}_{j} \left(\cM^{(3,x')}_I ; \mu^{(3)}_I\right)
  \ell(m^2_{j,I})\right) \quad {\bf (j)} \nonumber \\
&&  -2 \sum_{j\in\cM^{(3,x)}}
    \frac{\partial}{\partial m_{X,I}^2}
  \left(
 R^{[3,3]}_{j} \left(\cM^{(3,x)}_I ; \mu^{(3)}_I\right)
  \ell(m^2_{j,I})\right)  \quad {\bf (n)} \nonumber \\
&&  -2 \sum_{j\in\cM^{(3,y)}}
    \frac{\partial}{\partial m_{Y,I}^2}
  \left(
  R^{[3,3]}_{j} \left(\cM^{(3,y)}_I ; \mu^{(3)}_I\right)
  \ell(m^2_{j,I})\right)  \quad {\bf (o)} \nonumber \\
&& - \sum_{j\in\cM^{(4,x,x')}}
  R^{[4,3]}_{j} \left(\cM^{(4,x,x')}_I ; \mu^{(3)}_I\right)
  \ell(m^2_{j,I})\quad {\bf (l)} \nonumber\\
&& - \sum_{j\in\cM^{(4,x'y)}}
  R^{[4,3]}_{j} \left(\cM^{(4,x'y)}_I ; \mu^{(3)}_I\right)
  \ell(m^2_{j,I})\quad {\bf (k)} \nonumber\\
&& - 3\sum_{j\in\cM^{(4,x,y)}}
  R^{[4,3]}_{j} \left(\cM^{(4,x,y)}_I ; \mu^{(3)}_I\right)
  \ell(m^2_{j,I})\quad {\bf (m)} \Biggr]\nonumber\\
 + a^2\delta_V \,\Biggl[ && -\sum_{j\in\cM^{(4,x')}}
    \frac{\partial}{\partial m_{X',V}^2}
  \left(
  R^{[4,3]}_{j} \left(\cM^{(4,x')}_V ; \mu^{(3)}_V\right)
  \ell(m^2_{j,V})\right) \quad {\bf (j)} \nonumber \\
&&  - 2 \sum_{j\in\cM^{(4,x)}}
    \frac{\partial}{\partial m_{X,V}^2}
  \left(
 R^{[4,3]}_{j} \left(\cM^{(4,x)}_V ; \mu^{(3)}_V\right)
  \ell(m^2_{j,V})\right)  \quad {\bf (n)} \nonumber \\
&&  - 2 \sum_{j\in\cM^{(4,y)}}
    \frac{\partial}{\partial m_{Y,V}^2}
  \left(
 R^{[4,3]}_{j} \left(\cM^{(4,y)}_V ; \mu^{(3)}_V\right)
  \ell(m^2_{j,V})\right)  \quad {\bf (o)} \nonumber \\
&& + \sum_{j\in\cM^{(5,x,x')}}
  R^{[5,3]}_{j} \left(\cM^{(5,x,x')}_V ; \mu^{(3)}_V\right)
  \ell(m^2_{j,V})\quad {\bf (l)} \nonumber\\
&& + \sum_{j\in\cM^{(5,x'y)}}
  R^{[5,3]}_{j} \left(\cM^{(5,x'y)}_V ; \mu^{(3)}_V\right)
  \ell(m^2_{j,V})\quad {\bf (k)} \nonumber\\
&& - 3\sum_{j\in\cM^{(5,x,y)}}
  R^{[5,3]}_{j} \left(\cM^{(5,x,y)}_V ; \mu^{(3)}_V\right)
  \ell(m^2_{j,V})\quad {\bf (m)} \Biggr] + \Bigl[ V \to A \Bigr]\Biggr\}\ . \eqn{current-vertex-111}
\end{eqnarray}

\subsection{Results for $f_2(q^2)$}

Adding together \eqs{b22-compact}{current-vertex-compact} and $(Z_{P_{xx'}}+Z_{P_{yx'}})/2$ from \eq{ChPTWFR}, the complete one-loop result for the vector form factor in compact
notation is
\begin{eqnarray}
  f_2(q^2) = -\frac{1}{2(4\pi f)^2}\sum_\Xi &&\hspace{-3mm}\Biggl\{
  \frac{1}{16}
\sum_{\mathscr{S}}\left[  
 \ell\left(m^2_{x\mathscr{S},\Xi}\right)+
  \ell\left(m^2_{y\mathscr{S},\Xi}\right)
  + 4 \tilde B_{22}(m^2_{x\mathscr{S},\Xi},m^2_{y\mathscr{S},\Xi},q^2)\right]
  \nonumber\\&&{}
  +\frac{1}{4} \left[ \ell(D^\Xi_{XX}) + \ell(D^\Xi_{YY}) -2  \ell(D^\Xi_{XY})\right]
  \nonumber \\* &&{}
   \phantom{\frac{1}{4}\hspace{-3mm}}
   +\tilde B_{22}(m_{xy,\Xi}^2,D^\Xi_{XX},q^2) 
   +\tilde B_{22}(m_{xy,\Xi}^2,D^\Xi_{YY},q^2)
    \nonumber \\* &&{}   
    -2\tilde B_{22}(m_{xy,\Xi}^2,D^\Xi_{XY},q^2)
   \Biggr\} \ , \eqn{f2tot-compact}
\end{eqnarray}
where $\Xi$ runs over the sixteen independent meson tastes and $\mathscr{S}$ runs
over the three sea quark flavors. 
Note that the answer is independent of the spectator quark mass $m_{x'}$ at this order. The $m_{x'}$ 
dependence of the current vertex contribution cancels 
corresponding contributions from the wave function renormalization terms. This
appears to be necessary in order to satisfy the AG theorem, which is a statement about the
dependence on the valence quark masses $m_x$ and $m_y$.  Indeed, it is not hard to check that
\eq{f2tot-compact} obeys the AG theorem: As functions of the disconnected propagators, 
$\ell$ and $\tilde B_{22}$ are linear. Combining $D^\Xi_{XX}+D^\Xi_{YY}-2D^\Xi_{XY}$ using
\eq{DiscXi}, one easily extracts an overall factor of $(m_{Y,\Xi}^2-m_{X,\Xi}^2)^2 
\propto (m_y-m_x)^2$ from the disconnected terms (the taste splittings cancel in this 
difference). For the connected terms, those that are summed over $\mathscr{S}$, we may 
use \eq{B22expansion} to show that the contribution for each $\mathscr{S}$ is proportional 
to $(m^2_{y\mathscr{S}}-m^2_{x\mathscr{S}})^2 \propto (m_y-m_x)^2$ as $m_y\to m_x$.  
Thus $f_2(0)$ is second order in $m_y-m_x$ as required by the AG theorem.

We note also that \eq{f2tot-compact} is independent of the taste of the external mesons;
the taste-dependent factors $c_\Xi$ in \eqs{ChPTWFR}{current-vertex-compact} have canceled.  
Recall that we have taken the weak current to be a taste singlet (see \eq{Vmu}).  With a 
nonsinglet taste structure for the current, the result would of course depend on the taste 
of the mesons.

With the sea quark masses $m_u$, $m_d$, and $m_s$ nondegenerate (the 1+1+1 case), the 
result \eq{f2tot-compact} becomes, explicitly,
\begin{eqnarray}
  f_2^{N_f=1+1+1}(q^2) & =&-\frac{1}{2(4\pi f)^2}\Biggl\{
 \frac{1}{16} \sum_{\mathscr{S},\Xi}\left[
 \ell\left(m^2_{x\mathscr{S},\Xi}\right)+
  \ell\left(m^2_{y\mathscr{S},\Xi}\right)
  +4\, \tilde B_{22}(m^2_{x\mathscr{S},\Xi},m^2_{y\mathscr{S},\Xi},q^2)\right]
  \nonumber\\&&{}
  +\frac{1}{3}\Biggl[
  \sum_j \frac{\partial}{\partial m_{X,I}^2}
  \left(
  R^{[3,3]}_{j} \left(\cM^{(3,x)}_I ; \mu^{(3)}_I\right)
  \ell(m^2_{j,I})\right)\nonumber \\* &&{}
  +\sum_j \frac{\partial}{\partial m_{Y,I}^2}
    \left(
    R^{[3,3]}_{j}  \left(\cM^{(3,y)}_I ; \mu^{(3)}_I\right)
    \ell(m^2_{j,I})\right)    +2  \sum_{j}
  R^{[4,3]}_{j} \left(\cM^{(4,x,y)}_I ; \mu^{(3)}_I\right)
  \ell(m^2_{j,I})\nonumber\\* &&{}
  + 4 \frac{\partial}{\partial m_{X,I}^2} \left(
\sum_j \tilde B_{22}(m_{xy,I}^2,m_{j,I}^2,q^2) 
R^{[3,3]}_{j} 
\left(\cMI^{(3,x)} ; \mu^{(3)}_I\right)\right)\ \nonumber\\ 
&& + 4 \frac{\partial}{\partial m_{Y,I}^2} \left(
\sum_j \tilde B_{22}(m_{xy,I}^2,m_{j,I}^2,q^2) 
R^{[3,3]}_{j} 
\left(\cMI^{(3,y)} ; \mu^{(3)}_I\right)\right)
 \nonumber\\ 
&&+8 \sum_{j}
\tilde B_{22}(m_{xy,I}^2,m_{j,I}^2,q^2) R^{[4,3]}_j \left(\cMI^{(4,x,y)} ; 
\mu^{(3)}_I\right)  \Biggr]
  \nonumber \\* &&{}
+ a^2 \delta_V
  \Biggl[ 
    \sum_j
    \frac{\partial}{\partial m_{X,V}^2}
  \left(
  R^{[4,3]}_{j}  \left(\cM^{(4,x)}_V ; \mu^{(3)}_V\right)
  \ell(m^2_{j,V})
  \right)\nonumber \\* &&
  + \sum_j
    \frac{\partial}{\partial m_{Y,V}^2}
    \left(
    R^{[4,3]}_{j}\left(\cM^{(4,y)}_V ; \mu^{(3)}_V\right)
    \ell(m^2_{j,V})\right)
   +2\sum_{j}
  R^{[5,3]}_{j} \left(\cM^{(5,x,y)}_V ; \mu^{(3)}_V\right)
  \ell(m^2_{j,V})  \nonumber\\* &&{}  
  + 4 \frac{\partial}{\partial m_{X,V}^2}
\left( 
\sum_j \tilde B_{22}(m_{xy,V}^2,m_{j,V}^2,q^2) 
R^{[4,3]}_{j} \left(\cMV^{(4,x)} 
; \mu^{(3)}_V\right) \right) \nonumber\\ 
&& + 4 \frac{\partial}{\partial m_{Y,V}^2}
\left(
\sum_j \tilde B_{22}(m_{xy,V}^2,m_{j,V}^2,q^2) 
R^{[4,3]}_{j} \left(\cMV^{(4,y)} 
; \mu^{(3)}_V\right) \right) \nonumber\\ 
&&+8  \sum_{j}\, 
\tilde B_{22}(m_{xy,V}^2,m_{j,V}^2,q^2) R^{[5,3]}_j \left(\cMV^{(5,x,y)} ; 
\mu^{(3)}_V \right)\Biggl\rbrack +  \Bigl[ V \to A \Bigr]  \Biggr\} \ . \eqn{f2tot}
\end{eqnarray}

Keeping the valence masses arbitrary, but assuming exact isospin in the sea 
($m_u=m_d$), we can also find the explicit $N_f=2+1$ expression given in  
Appendix~\ref{ref:appB}. The partially quenched continuum results for 
the $N_f=1+1+1$ and $N_f=2+1$ cases are also given in Appendix~\ref{ref:appB}.

\section{Results for $f_+(q^2)$ in the mixed-action case}
\label{sec:mixed-results}

The only explicit difference between the mixed-action case and the unmixed 
theory discussed in the previous section is that mixed-action disconnected propagator 
for vector- and axial-tastes
has the form given in \eq{mixed-DiscXi}, rather than \eq{DiscXi}.  Of course there are also
implicit differences in the values the meson masses  for various tastes: taste splittings for
the valence (HISQ), sea (asqtad), and mixed mesons are given by $\Delta_\Xi^{\rm HISQ}$, 
$\Delta_\Xi^{\rm asqtad}$ and $\Delta_\Xi^{\rm mix}$, respectively. With these caveats, one may
use the result given in \eq{f2tot-compact} for the mixed case, just as for the unmixed
case.  More explicitly,  for $N_f=1+1+1$ in the mixed-action case we have 
\ba\eqn{1p1p1_f2tot_mixed}
f_2^{N_f=1+1+1}(q^2) & = & -\frac{1}{2(4\pi f)^2}\Biggl\{
  \frac{1}{16}
\sum_{\mathscr{S},\Xi}\left[ 
 \ell\left(m^2_{x\mathscr{S},\Xi}\right)+
  \ell\left(m^2_{y\mathscr{S},\Xi}\right)
+4\,
\tilde B_{22}(m_{x\mathscr{S},\Xi},m_{y\mathscr{S},\Xi},q^2)\right]
  \nonumber\\&&{}
  +\frac{1}{3}\Biggl[
  \sum_j \frac{\partial}{\partial m_{X,I}^2}
  \left(
  R^{[3,3]}_{j} \left(\cM^{(3,x)}_I ; \mu^{(3)}_I\right)
  \ell(m^2_{j,I})\right)\nonumber \\* &&{}
  +\sum_j \frac{\partial}{\partial m_{Y,I}^2}
    \left(
    R^{[3,3]}_{j}  \left(\cM^{(3,y)}_I ; \mu^{(3)}_I\right)
    \ell(m^2_{j,I})\right)    +2  \sum_{j}
  R^{[4,3]}_{j} \left(\cM^{(4,x,y)}_I ; \mu^{(3)}_I\right)
  \ell(m^2_{j,I})\nonumber\\* &&{}
  + 4 \frac{\partial}{\partial m_{X,I}^2} \left(
\sum_j \tilde B_{22}(m_{xy,I}^2,m_{j,I}^2,q^2)
R^{[3,3]}_{j}
\left(\cMI^{(3,x)} ; \mu^{(3)}_I\right)\right)\ \nonumber\\
&& + 4 \frac{\partial}{\partial m_{Y,I}^2} \left(
\sum_j \tilde B_{22}(m_{xy,I}^2,m_{j,I}^2,q^2)
R^{[3,3]}_{j}
\left(\cMI^{(3,y)} ; \mu^{(3)}_I\right)\right)
 \nonumber\\
&&+8 \sum_{j}
\tilde B_{22}(m_{xy,I}^2,m_{j,I}^2,q^2) R^{[4,3]}_j \left(\cMI^{(4,x,y)} ;
\mu^{(3)}_I\right)  \Biggr]
  \nonumber \\* &&{}
+  a^2 (\delta_V^{vv}-\delta_V^{\rm mix})
  \Biggl[
    \sum_j
    \frac{\partial}{\partial m_{X,V}^2}
  \left(
  R^{[4,3]}_{j}  \left(\cM^{(4,x)}_V ; \mu^{(3)}_V\right)
  \ell(m^2_{j,V})
  \right)\nonumber \\* &&
  + \sum_j
    \frac{\partial}{\partial m_{Y,V}^2}
    \left(
    R^{[4,3]}_{j}\left(\cM^{(4,y)}_V ; \mu^{(3)}_V\right)
    \ell(m^2_{j,V})\right)
   +2\sum_{j}
  R^{[5,3]}_{j} \left(\cM^{(5,x,y)}_V ; \mu^{(3)}_V\right)
  \ell(m^2_{j,V})  \nonumber\\* &&{}
  + 4 \frac{\partial}{\partial m_{X,V}^2}
\left\lbrace
\sum_j \tilde B_{22}(m_{xy,V}^2,m_{j,V}^2,q^2)
R^{[4,3]}_{j} \left(\cMV^{(4,x)}
; \mu^{(3)}_V\right) \right\rbrace \nonumber\\
&& + 4 \frac{\partial}{\partial m_{Y,V}^2}
\left\lbrace
\sum_j \tilde B_{22}(m_{xy,V}^2,m_{j,V}^2,q^2)
R^{[4,3]}_{j} \left(\cMV^{(4,y)}
; \mu^{(3)}_V\right) \right\rbrace \nonumber\\
&&+8  \sum_{j}\,
\tilde B_{22}(m_{xy,V}^2,m_{j,V}^2,q^2) R^{[5,3]}_j \left(\cMV^{(5,x,y)} ;
\mu^{(3)}_V \right)\Biggl\rbrack \nonumber\\
&&+ a^2\delta_V^{mix}
\Biggl\lbrack\frac{\partial \ell(m^2_{X,V}) }{\partial m_{X,V}^2}
+ \frac{\partial \ell(m^2_{Y,V}) }{\partial m_{Y,V}^2}
+2\; \frac{\left(\ell(m^2_{X,V}) -\ell(m^2_{Y,V})\right)}{m_{Y,V}^2-m_{X,V}^2}
\nonumber\\*
&&+  4\; \frac{\partial}{\partial m_{X,V}^2}\tilde B_{22}(m_{xy,V}^2,m_{X,V}^2,q^2)
+4\; \frac{\partial}{\partial m_{Y,V}^2}\tilde B_{22}(m_{xy,V}^2,m_{Y,V}^2,q^2)
\nonumber\\
&&+\;\frac{8}{m_{Y,V}^2-m_{X,V}^2}\left(\tilde B_{22}(m_{xy,V}^2,m_{X,V}^2,q^2)
-\tilde B_{22}(m_{xy,V}^2,m_{Y,V}^2,q^2)\right)
\Biggr\rbrack\nonumber\\
&&+  \Bigl[ V \to A \Bigr]  \Biggr\} \ ,
\ea
where again $\Xi$ runs over the sixteen independent meson tastes and $\mathscr{S}$ runs
over sea quark flavors. Since the Ademollo-Gatto theorem is obeyed at the level 
of \eq{f2tot-compact}, this mixed-action version must also obey the theorem.  
It is also easy to see directly that the terms proportional  to $\delta_{A,V}^{mix}$, 
which contain the mixed-action effects, vanish as  $(m_y-m_x)^2$ as $m_y\to m_x$.

If we take the isospin limit ($m_u=m_d$) of the expression above, 
the $N_f=2+1$ case, we obtain the result quoted in Eq.~(\ref{eq:2p1_f2tot_mixed}) of 
Appendix~\ref{ref:appB}. That result has already been given  in 
Ref.~\cite{Bazavov:2012cd} for the specific case $q^2=0$.

\section{Conclusions}
\label{sec:conclusion}

We have calculated the vector form factor $f_+(q^2)$ at one-loop in partially quenched S$\chi$PT for 
$N_f=1+1+1$, as well as in the isospin limit, $N_f=2+1$. We incorporate 
staggered effects for both the case where the action in the sea and the valence sectors are 
different (mixed-action) and the case in which all quarks are described with the same staggered 
action. We have found that, at this order, the form factor is independent of the spectator mass quark 
for any value of the momentum transfer. We also confirm that all our expressions obey the 
Ademollo-Gatto theorem.

The $N_f=2+1$ mixed-action expression in (\ref{eq:2p1_f2tot_mixed}) was used in the 
determination of $\vert V_{us}\vert$ by the Fermilab Lattice/MILC Collaboration in 
Ref.~\cite{Bazavov:2012cd}, and the $N_f=2+1$ unmixed expresion in (\ref{2p1_f2tot}) is used 
in the recent analysis by the same collaboration in Ref.~\cite{KtopiHISQ}. In this second work, 
the Fermilab Lattice/MILC Collaboration provides the most precise determination 
of $f_+(0)$, and the first one including simulations at the physical light quark 
masses. 

Even with simulations at the physical light quark masses, $\chi$PT is still very useful in miminizing 
the errors. The $\chi$PT formulation allows one to incorporate (more precise) data at heavier 
masses and correct for mistunings of the quark masses in the simulations, both in the valence 
and in the sea sector. It is also the perfect framework to incorporate analytically and 
systematically the corrections associated to the lattice artifacts, such as discretization 
and finite volume effects. This can be done for a specific lattice fermion formulation, as 
we have done here for staggered actions. Finite volume effects are now one of the dominant 
sources of error~\cite{KtopiHISQ}, so it is crucial to include those corrections in 
our PQS$\chi$PT formulae~\cite{Ktpi_SCHPT_FV} in order to achieve the 0.2\% precision in $f_+(0)$ 
required by the size of the experimental uncertainties.

\begin{acknowledgments}

This work was supported in part by the U.S. Department of Energy under Grants
No.~DE-FG02-91ER40628 (C.B.), by
the European Community SP4-Capacities
``Study of Strongly Interacting Matter''
(HadronPhysics3, Grant Agreement n.\ 283286),
the Swedish Research Council  grants 621-2011-5080 and 621-2010-3326 (J.B.); 
by the MINECO (Spain) under grant FPA2010-16696 and Ram\'on y Cajal program (E.G.);
by the Junta de Andaluc\'ia (Spain) under Grants No.~FQM-101 and 
No.~FQM-6552 (E.G.); 
and by European Commission (EC) under Grant No.~PCIG10-GA-2011-303781 (E.G.).
C.B.\ thanks the Galileo
Galilei Institute for Theoretical Physics for the hospitality
and the INFN for partial support while this work was in progress, and Maarten Golterman for discussions.

\end{acknowledgments}

\appendix
\section{One-loop integrals}
\label{app:integrals}

We need the following two Euclidean integrals
\begin{eqnarray}
\cA(m^2) &\equiv &\int \frac {d^4 k}{(2\pi)^4}\; \frac{1}{k^2+m^2} \ , \eqn{A-def}\\ 
\cB_{\mu\nu}(m_1^2,m_2^2,q^2) &\equiv &\int \frac {d^4 k}{(2\pi)^4}\; \frac{k_\mu k_\nu}{(k^2+m_1^2)
((k-q)^2+m_2^2)} \ ,\eqn{Bmunu-def} \\
&=& q_\mu q_\nu \cC_{21}(m_1^2,m_2^2,q^2) - \delta_{\mu\nu} \cC_{22}(m_1^2,m_2^2,q^2)\eqn{C22-def}
\end{eqnarray}
After regularization and renormalization following \rcite{Bijnens:2002hp}, one has
\begin{eqnarray}
\cA(m^2) &\to &\frac {1}{16\pi^2}\; \ell(m^2) \ , \eqn{Avalue}\\ 
\cC_{22}(m_1^2,m_2^2,q^2) &\to &\frac {1}{16\pi^2}\;\tilde B_{22}(m_1^2,m_2^2,-q^2) \eqn{Cvalue}
\end{eqnarray}
The chiral logarithm function $\ell$ is given by
\begin{equation}\label{eq:ell}
\ell(m^2)\equiv m^2\ln(m^2/\Lambda_\chi^2)\ ,
\end{equation} 
with $\Lambda_\chi$ the chiral scale. The function
$\tilde B_{22}(m_1^2, m_2^2, s)$ is related to the function $\bar B_{22}$
defined in Ref.~\cite{Bijnens:2002hp} by
\begin{equation}\label{eq:B22}
\tilde B_{22}(m_1^2, m_2^2, s) = (4\pi)^2\;\bar B_{22}(m_1^2, m_2^2, s, \Lambda_\chi^2)\ .
\end{equation}
The minus sign for the $q^2$ argument in \eq{Cvalue} arises because $q$ in \eq{Bmunu-def} is defined to be a Euclidean momentum; the physical momentum transfer squared is $s=-q^2$.
Explicitly, $\tilde B_{22}$ is given by
\begin{eqnarray}
\tilde B_{22}(m_1^2, m_2^2, s) &=&\frac{1}{6}\left[-\ell(m_2^2) + 2m_1^2\tilde B(m_1^2, m_2^2, s) -
(s+m_1^2-m_2^2)\tilde B_1(m_1^2, m_2^2, s) \right] \nonumber \\
&&+\frac{1}{18}\left[3m_1^2+3m_2^2-s\right]\ ,\eqn{B22-value}
\end{eqnarray}
where
\begin{eqnarray}
\tilde B(m_1^2, m_2^2, s) &=&
\frac{1}{2}\left[2+\left(\frac{\Sigma}{\Delta}-\frac{\Delta}{s}\right)
\ln\frac{m_1^2}{m_2^2} -\frac{\nu}{s}\ln\frac{(s+\nu)^2-\Delta^2}{(s-\nu)^2-\Delta^2}\right] \nonumber\\
&&-\frac{\ell(m_1^2)-\ell(m_2^2)}{m_1^2-m_2^2} \ , 
\eqn{B-value}\\
\tilde B_1(m_1^2, m_2^2, s) &=&\frac{1}{2s}\left(\ell(m_1^2)-\ell(m_2^2)
+(m_1^2-m_2^2+s)\tilde B(m_1^2, m_2^2, s)\right)\ ,\eqn{B1-value}
\end{eqnarray}
with $\Delta = m_1^2- m_2^2$, $\Sigma=m_1^2+ m_2^2$, and
$\nu^2 = [s-(m_1+m_2)^2][s-(m_1-m_2)^2]$.

In the special case of $s=q^2=0$, $\tilde B_{22}$ takes the simple form
\begin{equation}\label{eq:B22q0}
\tilde B_{22}(m_1^2, m_2^2, 0) = -\frac{1}{4}\left(\frac{
m_2^2\,\ell(m_2^2) - m_1^2\,\ell(m_1^2)}{m_2^2-m_1^2}\right)+\frac{1}{8}\left(m_1^2
+ m_2^2\right) \ .
\end{equation}
For checking the AG theorem, we need the behavior of $\tilde B_{22}(m_1^2, m_2^2, 0) $ as
$m^2_2\to m^2_1$.  Letting $m^2_1 =  m^2$, $m^2_2 =  m^2 + \epsilon$, and expanding \eq{B22q0}
through ${\cal O}(\epsilon)$, we find
\begin{equation}\label{eq:B22expansion}
\tilde B_{22}(m^2, m^2+\epsilon, 0) = -\frac{1}{4}\Big[\ln(m^2/\Lambda^2_\chi)(2m^2+\epsilon) +\epsilon\Big] +{\cal O}(\epsilon^2)\ .
\end{equation}


For diagrams with neutral particles in the loop, the simple propagator in \eq{A-def}, or one of the
two propagators in \eq{Bmunu-def}, may be replaced by a disconnected propagator, \eq{DiscXi} or 
\eq{DiscI}.  For an explicit representation of the result of the integrals, one may follow 
the standard procedure and write the disconnected propagator as a sum over residues times simple poles, and apply \eqs{Avalue}{Cvalue}. 
However, this produces complicated expressions that depend on the details of the
sea sector (\eg are different in the 1+1+1 case and the 2+1 case, as well in the mixed-action and unmixed
cases).  To see the overall structure of the
results more clearly and compactly, it is useful also to have a notation that writes the 
integrals as functions of the disconnected propagator itself.  Thus we allow a replacement of the argument
$m^2$ in \eq{A-def}
or $m_2^2$ in \eq{Bmunu-def} by a disconnected propagator and define
\begin{eqnarray}
\cA(D^\Xi) &\equiv &\int \frac {d^4 k}{(2\pi)^4}\; D^\Xi(k) \ , \eqn{Adisc-def}\\ 
\cB_{\mu\nu}(m_1^2,D^\Xi,q^2) &\equiv &\int \frac {d^4 k}{(2\pi)^4}\; \frac{k_\mu k_\nu}{(k^2+m_1^2)}
D^\Xi(k-q)\ ,\eqn{Bmunudisc-def} \\
&=& q_\mu q_\nu \cC_{21}(m_1^2,D^\Xi,q^2) - \delta_{\mu\nu} \cC_{22}(m_1^2,D^\Xi,q^2)
\end{eqnarray}
The corresponding expressions after regularization and renormalization are then denoted as 
\begin{eqnarray}
\cA(D^\Xi) &\to &\frac {1}{16\pi^2}\; \ell(D^\Xi) \ , \eqn{Adisc-value}\\ 
\cC_{22}(m_1^2,D^\Xi,q^2) &\to &\frac {1}{16\pi^2}\;\tilde B_{22}(m_1^2,D^\Xi,-q^2) \eqn{Cdisc-value}\ .
\end{eqnarray}

\section{Form factor in the isospin limit $N_f=2+1$ and in the 
continuum}

\label{ref:appB}

In this Appendix we collect the isospin limit and the continuum limit of the 
one-loop PQS\chpt\, in \eq{f2tot} and \eq{1p1p1_f2tot_mixed}.

\subsection{Unmixed case}

The vector form factor at one loop in PQS\chpt\, and in the isospin limit, 
$N_f=2+1$ is
\begin{eqnarray}
  f_2^{N_f=2+1}(q^2) & =&-\frac{1}{2(4\pi f)^2}\Biggl\{
  \frac{1}{16}\sum_{\mathscr{S},\Xi}\left[
 \ell\left(m^2_{x\mathscr{S},\Xi}\right)+
  \ell\left(m^2_{y\mathscr{S},\Xi}\right)
 +4\, 
\tilde B_{22}(m_{x\mathscr{S},\Xi},m_{y\mathscr{S},\Xi},q^2)\right]
  \nonumber\\&&{}
  +\frac{1}{3}\Biggl[
  \sum_j \frac{\partial}{\partial m_{X,I}^2}
  \left(
  R^{[2,2]}_{j} \left(\cM^{(2,x)}_I ; \mu^{(2)}_I\right)
  \ell(m^2_{j,I})\right)\nonumber \\* &&{}
  +\sum_j \frac{\partial}{\partial m_{Y,I}^2}
    \left(
    R^{[2,2]}_{j}  \left(\cM^{(2,y)}_I ; \mu^{(2)}_I\right)
    \ell(m^2_{j,I})\right)    +2  \sum_{j}
  R^{[3,2]}_{j} \left(\cM^{(3,x,y)}_I ; \mu^{(2)}_I\right)
  \ell(m^2_{j,I})\nonumber\\* &&{}
  + 4 \frac{\partial}{\partial m_{X,I}^2} \left(
\sum_j \tilde B_{22}(m_{xy,I}^2,m_{j,I}^2,q^2) 
R^{[2,2]}_{j} 
\left(\cMI^{(2,x)} ; \mu^{(2)}_I\right)\right)\ \nonumber\\ 
&& + 4 \frac{\partial}{\partial m_{Y,I}^2} \left(
\sum_j \tilde B_{22}(m_{xy,I}^2,m_{j,I}^2,q^2) 
R^{[2,2]}_{j} 
\left(\cMI^{(2,y)} ; \mu^{(2)}_I\right)\right)
 \nonumber\\ 
&&+8 \sum_{j}
\tilde B_{22}(m_{xy,I}^2,m_{j,I}^2,q^2) R^{[3,2]}_j \left(\cMI^{(3,x,y)} ; 
\mu^{(2)}_I\right)  \Biggr]
  \nonumber \\* &&{}
 + a^2 \delta_V
  \Biggl[ 
    \sum_j
    \frac{\partial}{\partial m_{X,V}^2}
  \left(
  R^{[3,2]}_{j}  \left(\cM^{(3,x)}_V ; \mu^{(2)}_V\right)
  \ell(m^2_{j,V})
  \right)\nonumber \\* &&
  + \sum_j
    \frac{\partial}{\partial m_{Y,V}^2}
    \left(
    R^{[3,2]}_{j}\left(\cM^{(3,y)}_V ; \mu^{(2)}_V\right)
    \ell(m^2_{j,V})\right)
   +2\sum_{j}
  R^{[4,2]}_{j} \left(\cM^{(4,x,y)}_V ; \mu^{(2)}_V\right)
  \ell(m^2_{j,V})  \nonumber\\* &&{}  
  + 4 \frac{\partial}{\partial m_{X,V}^2}
\left( 
\sum_j \tilde B_{22}(m_{xy,V}^2,m_{j,V}^2,q^2) 
R^{[3,2]}_{j} \left(\cMV^{(3,x)} 
; \mu^{(2)}_V\right) \right) \nonumber\\ 
&& + 4 \frac{\partial}{\partial m_{Y,V}^2}
\left( 
\sum_j \tilde B_{22}(m_{xy,V}^2,m_{j,V}^2,q^2) 
R^{[3,2]}_{j} \left(\cMV^{(3,y)} 
; \mu^{(2)}_V\right) \right) \nonumber\\ 
&&+8  \sum_{j}\, 
\tilde B_{22}(m_{xy,V}^2,m_{j,V}^2,q^2) R^{[4,2]}_j \left(\cMV^{(4,x,y)} ; 
\mu^{(2)}_V \right)\Biggl\rbrack +  \Bigl[ V \to A \Bigr]  \Biggr\} \ , \label{2p1_f2tot}
\end{eqnarray}
where the mass sets $\cM^{(2,x)},\ \cM^{(3,x,y)},\ \dots$ correspond to those of
\eq{denom_mass_sets} but with the $\pi^0$ mass eliminated and the first subscript reduced by 1.
Similarly, the set $\mu^{(2)}$ corresponds to $\mu^{(3)}$ in \eq{denom_mass_sets}, with the $D$ mass eliminated.

In the continuum, the partially quenched $N_f=1+1+1$ result in \eq{f2tot} reduces to
\ba
f_2^{N_f=1+1+1}(q^2) & =&-\frac{1}{2(4\pi f)^2}\Biggl\{
  \sum_{\mathscr{S}}\left[
 \ell\left(m^2_{x\mathscr{S}}\right)+
  \ell\left(m^2_{y\mathscr{S}}
  \right) + 4\tilde B_{22}(m_{x\mathscr{S}},m_{y\mathscr{S}},q^2)\right]
  \nonumber\\&&{}
  +\frac{1}{3}\Biggl[
  \sum_j \frac{\partial}{\partial m_{X}^2}
  \left(
  R^{[3,3]}_{j}\left(\cM^{(3,x)} ; \mu^{(3)}\right)
  \ell(m^2_{j})\right)\nonumber \\* &&{}
  +\sum_j \frac{\partial}{\partial m_{Y}^2}
    \left(
    R^{[3,3]}_{j}  \left(\cM^{(3,y)} ; \mu^{(3)}\right)
    \ell(m^2_{j})\right)    +2  \sum_{j}
  R^{[4,3]}_{j} \left(\cM^{(4,x,y)} ; \mu^{(3)}\right)
  \ell(m^2_{j})\nonumber\\* &&{}
  + 4 \frac{\partial}{\partial m_{X}^2} \left(
\sum_j \tilde B_{22}(m_{xy}^2,m_{j}^2,q^2)
R^{[3,3]}_{j}
\left(\cM^{(3,x)} ; \mu^{(3)}\right)\right)\ \nonumber\\
&& + 4 \frac{\partial}{\partial m_{Y}^2} \left(
\sum_j \tilde B_{22}(m_{xy}^2,m_{j}^2,q^2)
R^{[3,3]}_{j} \left(\cM^{(3,y)} ; \mu^{(3)}\right)\right)
 \nonumber\\
&&+8 \sum_{j}
\tilde B_{22}(m_{xy}^2,m_{j}^2,q^2) R^{[4,3]}_j \left(\cM^{(4,x,y)} ;
\mu^{(3)}\right)  \Biggr]\Biggr\} \ , 
\ea
where the meson masses $m_{xy}^2$ are those in the continuum, the residue functions 
$R^{[n,k]}(\cM;\mu)$ are the same as in the staggered expressions, and $\cM$ and $\mu$ are the 
set of meson masses defined in Eq.~(\ref{eq:denom_mass_sets}) but in the continuum.   

In the partially quenched $N_f=2+1$ case, the result in Eq.~(\ref{2p1_f2tot})  reduces 
in the continuum to
\ba
f_2^{N_f=2+1}(q^2) & =& -\frac{1}{2(4\pi f)^2}\Biggl\{
  \sum_{\mathscr{S}}\left[
 \ell\left(m^2_{x\mathscr{S}}\right)+
  \ell\left(m^2_{y\mathscr{S}}\right) +
4\tilde B_{22}(m_{x\mathscr{S}},m_{y\mathscr{S}},q^2)\right]
  \nonumber\\&&{}
  +\frac{1}{3}\Biggl[
  \sum_j \frac{\partial}{\partial m_{X}^2}
  \left(
  R^{[2,2]}_{j} \left(\cM^{(2,x)} ; \mu^{(2)}\right)
  \ell(m^2_{j})\right)\nonumber \\* &&{}
  +\sum_j \frac{\partial}{\partial m_{Y}^2}
    \left(
    R^{[2,2]}_{j}  \left(\cM^{(2,y)} ; \mu^{(2)}\right)
    \ell(m^2_{j})\right)    +2  \sum_{j}
  R^{[3,2]}_{j} \left(\cM^{(3,x,y)} ; \mu^{(2)}\right)
  \ell(m^2_{j})\nonumber\\* &&{}
  + 4 \frac{\partial}{\partial m_{X}^2} \left(
\sum_j \tilde B_{22}(m_{xy}^2,m_{j}^2,q^2)
R^{[2,2]}_{j}
\left(\cM^{(2,x)} ; \mu^{(2)}\right)\right)\ \nonumber\\
&& + 4 \frac{\partial}{\partial m_{Y}^2} \left(
\sum_j \tilde B_{22}(m_{xy}^2,m_{j}^2,q^2)
R^{[2,2]}_{j}
\left(\cM^{(2,y)} ; \mu^{(2)}\right)\right)
 \nonumber\\
&&+8 \sum_{j}
\tilde B_{22}(m_{xy}^2,m_{j}^2,q^2) R^{[3,2]}_j \left(\cM^{(3,x,y)} ;
\mu^{(2)}\right)  \Biggr]\Biggr\} \ .
\label{continuumf2+1}
\ea
The continuum \chpt\, result $f_2^{N_f=2+1}$ for zero momentum transfer, $q^2=0$,
was already given in the Appendix of Ref.~\cite{Becirevic:2005py}, but we believe there 
was a misprint affecting the sign of the last term in second line of preprint version.

\subsection{Mixed-action case}

Taking the limit $m_u=m_d$ (the 2+1 case) of the mixed action result, 
\eq{1p1p1_f2tot_mixed}, we get
\begin{eqnarray}\label{eq:2p1_f2tot_mixed}
 f_2^{N_f=2+1}(q^2) & =&
\frac{-1}{2(4\pi f)^2}\Biggl\{
  \frac{1}{16}\sum_{\mathscr{S},\Xi}
  \left[
  \ell\left(m^2_{x\mathscr{S},\Xi}\right)
  +\ell\left(m^2_{y\mathscr{S},\Xi}\right)
  +4\tilde B_{22}(m_{x\mathscr{S}},m_{y\mathscr{S}},q^2)\right]\nonumber \\* &&
  +\frac{1}{3}\Biggl[\sum_{j}
    \frac{\partial}{\partial m_{X,I}^2}
  \left(
  R^{[2,2]}_{j} \left(\cM^{(2,x)}_I ; \mu^{(2)}_I\right)
  \ell(m^2_{j,I})\right)\nonumber \\* &&
  +\sum_{j}
    \frac{\partial}{\partial m_{Y,I}^2}
    \left(
    R^{[2,2]}_{j}  \left(\cM^{(2,y)}_I ; \mu^{(2)}_I\right)
    \ell(m^2_{j,I})\right) +2\sum_{j}
  R^{[3,2]}_{j} \left(\cM^{(3,x,y)}_I ; \mu^{(2)}_I\right)
  \ell(m^2_{j,I})\Biggr]
  \nonumber \\* &&
+ \frac{4}{3}\;\frac{\partial}{\partial m_{X,I}^2} \left(
\sum_j \tilde B_{22}(m_{xy,I}^2,m_{j,I}^2,q^2) 
R^{[2,2]}_{j} 
\left(\cMI^{(2,x)} ; \mu^{(2)}_I\right)\right)\
\nonumber\\ 
&& + \frac{4}{3}\;\frac{\partial}{\partial m_{Y,I}^2} \left( 
\sum_j \tilde B_{22}(m_{xy,I}^2,m_{j,I}^2,q^2) 
R^{[2,2]}_{j} 
\left(\cMI^{(2,y)} ; \mu^{(2)}_I\right)\right)\
\nonumber\\ 
&&+\frac{8}{3}\; \sum_{j}\, 
\tilde B_{22}(m_{xy,I}^2,m_{j,I}^2,q^2) R^{[3,2]}_j \left(\cMI^{(3,x,y)} ; 
\mu^{(2)}_I\right) \nonumber\\
&&+ a^2 (\delta_V^{vv}-\delta_V^{mix})
\Biggl[
    \sum_{j}
    \frac{\partial}{\partial m_{X,V}^2}
  \left(
  R^{[3,2]}_{j}  \left(\cM^{(3,x)}_V ; \mu^{(2)}_V\right)
  \ell(m^2_{j,V})
  \right)\nonumber \\* &&
  +\sum_{j}
    \frac{\partial}{\partial m_{Y,V}^2}
    \left(
    R^{[3,2]}_{j}\left(\cM^{(3,y)}_V ; \mu^{(2)}_V\right)
    \ell(m^2_{j,V})\right)\nonumber
  +2\sum_{j}
  R^{[4,2]}_{j}\left(\cM^{(4,x,y)}_V ; \mu^{(2)}_V\right)
  \ell(m^2_{j,V})
    \nonumber \\* 
&&  
+ 4\; \frac{\partial}{\partial m_{X,V}^2}
\left( 
\sum_j \tilde B_{22}(m_{xy,V}^2,m_{j,V}^2,q^2) 
R^{[3,2]}_{j} \left(\cMV^{(3,x)} 
; \mu^{(2)}_V\right) \right)
\\ 
&& +4\;\frac{\partial}{\partial m_{Y,V}^2}
\left( 
\sum_j \tilde B_{22}(m_{xy,V}^2,m_{j,V}^2,q^2) 
R^{[3,2]}_{j} \left(\cMV^{(3,y)} 
; \mu^{(2)}_V\right) \right)
 \nonumber\\ 
&&+8\; \sum_{j}\, 
\tilde B_{22}(m_{xy,V}^2,m_{j,V}^2,q^2) R^{[4,2]}_j \left(\cMV^{(4,x,y)} ; 
\mu^{(2)}_V\right) \,\Biggl\rbrack\nonumber\\
&&+ a^2\delta_V^{mix}
\Biggl\lbrack\frac{\partial \ell(m^2_{X,V}) }{\partial m_{X,V}^2}
+ \frac{\partial \ell(m^2_{Y,V}) }{\partial m_{Y,V}^2} 
+2\; \frac{\left(\ell(m^2_{X,V}) -\ell(m^2_{Y,V})\right)}{m_{Y,V}^2-m_{X,V}^2}
\nonumber\\*
&&+  4\; \frac{\partial}{\partial m_{X,V}^2}\tilde B_{22}(m_{xy,V}^2,m_{X,V}^2,q^2)
+4\; \frac{\partial}{\partial m_{Y,V}^2}\tilde B_{22}(m_{xy,V}^2,m_{Y,V}^2,q^2)
\nonumber\\
&&+\;\frac{8}{m_{Y,V}^2-m_{X,V}^2}\left(\tilde B_{22}(m_{xy,V}^2,m_{X,V}^2,q^2)
-\tilde B_{22}(m_{xy,V}^2,m_{Y,V}^2,q^2)\right)
\Biggr\rbrack\nonumber\\
&& +  \Bigl[ V \to A \Bigr]  \Biggl\}\ , \nonumber 
\end{eqnarray}

\end{document}